\documentclass[manuscript]{aastex6}

\fullcollaborationName{}
\begin{document}

\title{ALMA View of the Galactic Center Mini-spiral: Ionized Gas Flows around Sagittarius A$^\star$}

\author{Masato Tsuboi\altaffilmark{1} and Yoshimi Kitamura}
\affil{Institute of Space and Astronautical Science(ISAS), Japan Aerospace Exploration Agency,\\
3-1-1 Yoshinodai, Chuo-ku, Sagamihara, Kanagawa 252-5210, Japan}
\author{Kenta Uehara}
\affil{Department of Astronomy, The University of Tokyo, Bunkyo, Tokyo 113-0033, Japan}
\author{Ryosuke Miyawaki}
\affil{Colledge of Arts and Sciences, J.F. Oberlin University, Machida, Tokyo 194-0294, Japan}
\author{Takahiro Tsutsumi}
\affil{National Radio Astronomy Observatory,  Socorro, NM 87801-0387, USA}
\author{Atsushi Miyazaki}
\affil{Japan Space Forum, Kandasurugadai, Chiyoda-ku,Tokyo,101-0062 and National Astronomical Observatory of Japan, Mitaka, Tokyo 181-8588, Japan}
\author{Makoto Miyoshi}
\affil{National Astronomical Observatory of Japan, Mitaka, Tokyo 181-8588, Japan}
\altaffiltext{1}{tsuboi@vsop.isas.jaxa.jp}

\begin{abstract}
  We have performed the observation of the``Galactic Center Mini-spiral(GCMS)" in H42$\alpha$ recombination line as a part of  the first large-scale mosaic observation in the Sagittarius A complex using  Atacama Millimeter/sub-millimeter Array (ALMA). 
We revealed the kinematics of the ionized gas streamers of the GCMS.
We found that the ionized gas streamers of the Northern Arm(NA) and Eastern Arm(EA) in their outer regions somewhat deviate  from the Keplerian orbits which were derived previously from the trajectories in the inner regions. 
In addition, we found that the streamer corresponding to the Bar of the GCMS has a Keplerian orbit with an eccentricity of $e\sim0.8$, which is independent from the Keplerian orbits of the other streamers of the GCMS. 
We estimated the LTE electron temperature and  electron density in the ionized gas streamers.
We confirmed the previously claimed tendency that the electron temperatures increase toward Sgr A$^\ast$. 
We found that the electron density in the NA and EA also increases with approaching Sgr A* without the lateral expansion of the gas streamers. 
This suggests that there is some external pressure around the GCMS. The ambient ionized gas may cause the confinement and/or the perturbation for the orbits.
There is a good positional correlation between the protostar candidates detected by JVLA at 34 GHz and the ionized gas streamer, Northeastern Arm, newly found by our H42$\alpha$ recombination line observation. This suggests that the candidates had formed in the streamer and they were brought to near Sgr A* as the streamer falls.

\end{abstract}


\received{}
\accepted{}

\section{Introduction}

Sagittarius A$^\ast$ (Sgr A$^\ast$) is a compact source from radio to X-ray wavelengths associated with the Galactic center super massive black hole (GCBH), which is very close to the dynamical center of Milky Way Galaxy \citep{Reid} and has the mass of $\sim4\times10^6 $M$_\sun$ \citep[e.g.][]{Ghez,Gillessen}. 
A bundle of the ionized gas streams located within 2 pc of Sgr A$^\ast$ was identified as the ``Galactic Center Mini-spiral (GCMS)" using Very Large Array (VLA) and IR telescopes \citep[e.g.][]{Lacy1980,Ekers1983,LO1983,Scoville}.  The kinematic structure of the GCMS has been studied mainly by these telescopes \citep[e.g.][]{SerabynLacy, Serabyn1988, Lacy1991, Roberts, Zhao2009, Zhao2010}.  The stretched appearance and kinematics of the GCMS suggest the models in which the streamers  are tentative structures with Keplerian orbits around Sgr A$^\ast$ \citep[e.g.][]{Serabyn1988, Zhao2009, Zhao2010}. There are still alternate models to explain these properties \citep[e.g. one-armed spiral;][]{Lacy1991,Irons}. 
The tidal force of Sgr A$^\star$ must have a serious effect on the interstellar medium (ISM) in the vicinity of Sgr A$^\ast$, i.e., the GCMS \citep[e.g.][] {Lacy1982}. Furthermore, the strong Lyman continuum radiation from the OB and WR stars in the Central cluster ionizes the ISM rapidly  in the region \citep[e.g.][]{Genzel1996, Paumard}. 
A recent dust observation of the GCMS with ALMA suggested that there is a possible scenario for the formation of the Central cluster: a star-forming molecular cloud is falling from a region somewhat far from Sgr A$^\ast$ and supplies young stars and the ISM to the vicinity of Sgr A$^\ast$ \citep[e.g.][] {Tsuboi2016}.
However, it is not clearly demonstrated how the gas is fed to Sgr A$^\ast$ because the innermost part of the ionized gas has very weak intensity and complicated structures. 

We present new observational results on the  ionized gas streams in the vicinity of Sgr A$^\ast$ based on the observation of the H42$\alpha$ recombination line using the Atacama Large Millimeter/submillimeter Array (ALMA). 
The distance of the Galactic Center is assumed to be 8 kpc in this paper.

\section{Observation and Data Reduction}
We have performed the observation of the GCMS in H42$\alpha$ recombination line ($85.6884$ GHz) as a part of  ALMA Cy.1 observation (2012.1.00080.S. PI M.Tsuboi), which is an ionized gas  tracer.  The entire ALMA observation consists of a 137 pointing mosaic of the 12-m array and a 52 pointing mosaic of the 7-m array (ACA), covering a $330\arcsec \times 330\arcsec$ area including  the ``Galactic Center 50 km s$^{-1}$ Molecular Cloud" and  the GCMS  in CS $J=2-1$ ($97.980953$ GHz), SiO $v=0~J=2-1$ ($86.846995$ GHz), H$^{13}$CO$^+ J=1-0$ ($86.754288$ GHz) and H42$\alpha$ emission lines. 
The molecular cloud encompasses the most conspicuous star forming region in the vicinity of Sgr A$^\ast$. We will present  the detailed description of the analysis and the full results in another paper. We have detected the recombination line in the GCMS and the Sgr A East HII region, and  we concentrate on the GCMS in H42$\alpha$ recombination line in this paper.   The data of the recombination line have angular resolutions of $2.48\arcsec \times 1.86\arcsec, PA=89.3^\circ$ and $1.87\arcsec \times 1.37\arcsec, PA=84.0^\circ$ using ``natural weighting" and ``Briggs weighting (R=0.5)" as {\it u-v} sampling, respectively.    The frequency channel width is 244 kHz. The velocity resolution is $1.7$ km s$^{-1}$(488 kHz).   J0006-0623, J1517-2422, J717-3342,  J1733-1304, J1743-3058, J1744-3116 and J2148+0657 were used as phase calibrators. The flux density scale was determined using Titan, Neptune and Mars. Because the observation has a large time span of one year and seven months,  the absolute flux density accuracy  may be as large as 15 \%. The calibration and imaging of the data were done by CASA \citep{McMullin}. The continuum emission of  the GCMS and Sgr A$^\ast$  was subtracted from the spectral data using the CASA task UVCONTSUB ($fitorder=1$). 
Although the flux density of Sgr A* at 100 GHz varied in the range of 1 - 2 Jy, the residual emission seen at the position of Sgr A$^{\ast}$ in the channel maps is as small as 5-10 mJy beam$^{-1}$ (see Figures 2 and 3). This suggests that the contamination from the continuum emission of  the GCMS is at most $\sim1$ \%.

\begin{figure}
\figurenum{1}
\begin{center}
\includegraphics[width=12cm]{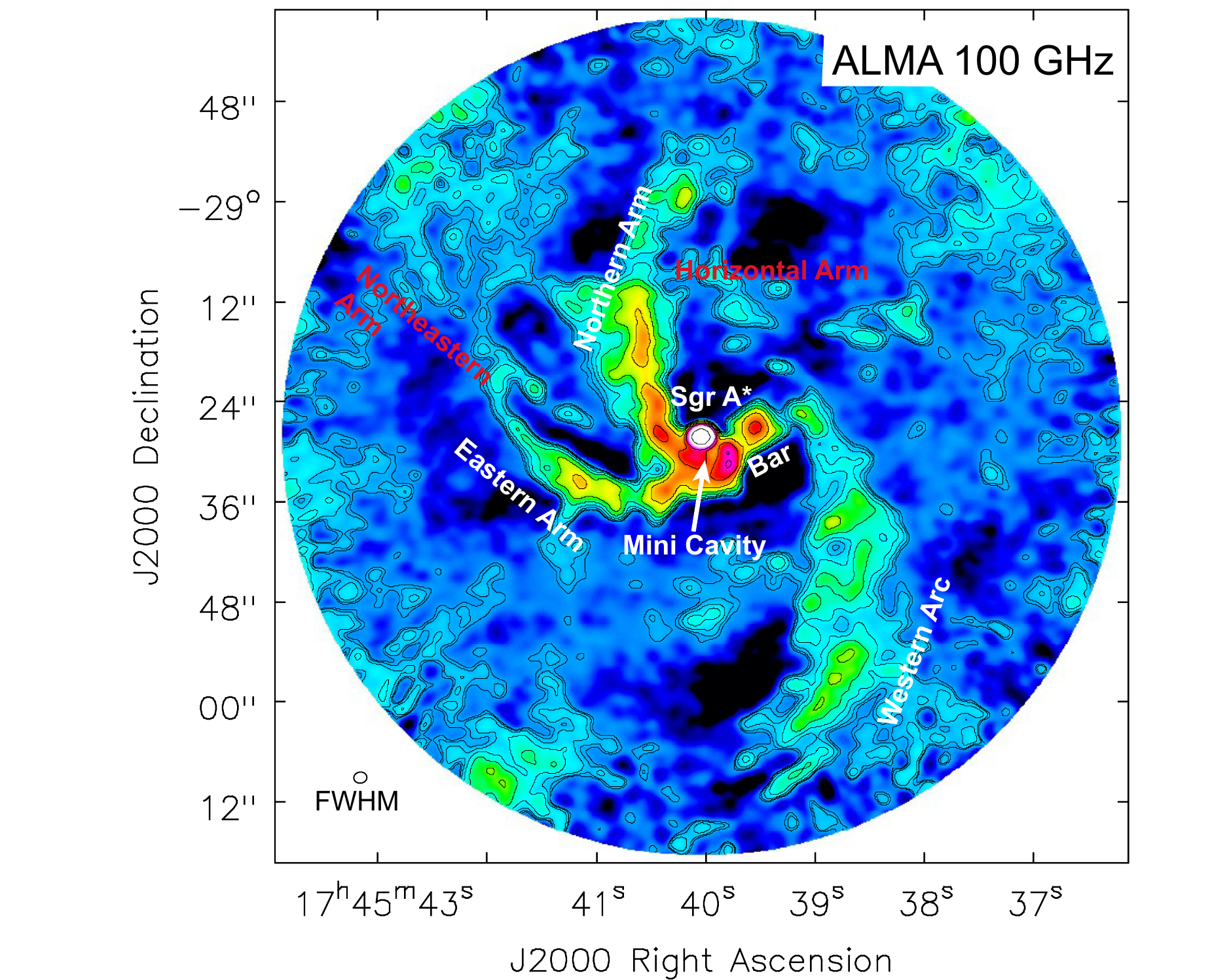}
\caption{ALMA view of the ``Galactic Center Mini-spiral (GCMS)" at 100 GHz \citep{Tsuboi2016}. This is a finding chart of the substructures of the GCMS. The correction for primary beam attenuation  is applied in the figure. }
\end{center}
\end{figure}

\begin{figure}
\figurenum{2}
\includegraphics[width=14cm]{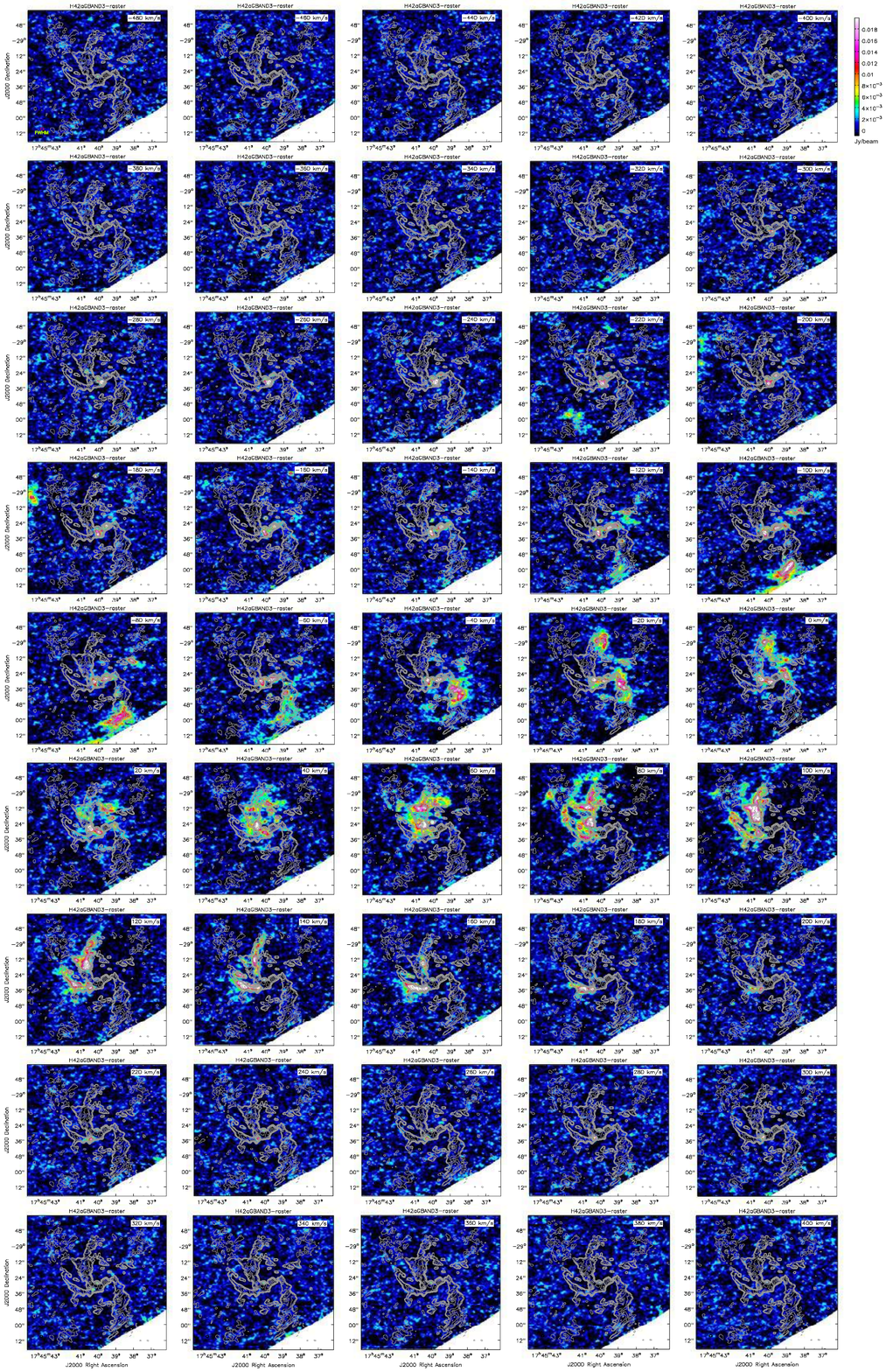}
\caption{Channel maps of the ``Galactic Center Mini-spiral (GCMS)"  in the H42$\alpha$ recombination line (pseudo color). The angular resolution is $2.5\arcsec \times 1.9\arcsec$($PA=90^\circ$) using ``natural weighting", which is shown on the lower-left corner 
 as an open oval.  The central velocity is shown in the upper right corner . The velocity width is $\Delta V=-20$ km s$^{-1}$. The contours in the figure show the continuum emission of the GCMS at 100 GHz for comparison \citep{Tsuboi2016}. }
\end{figure}

\begin{figure}
\figurenum{3}
\begin{center}
\includegraphics[width=18cm]{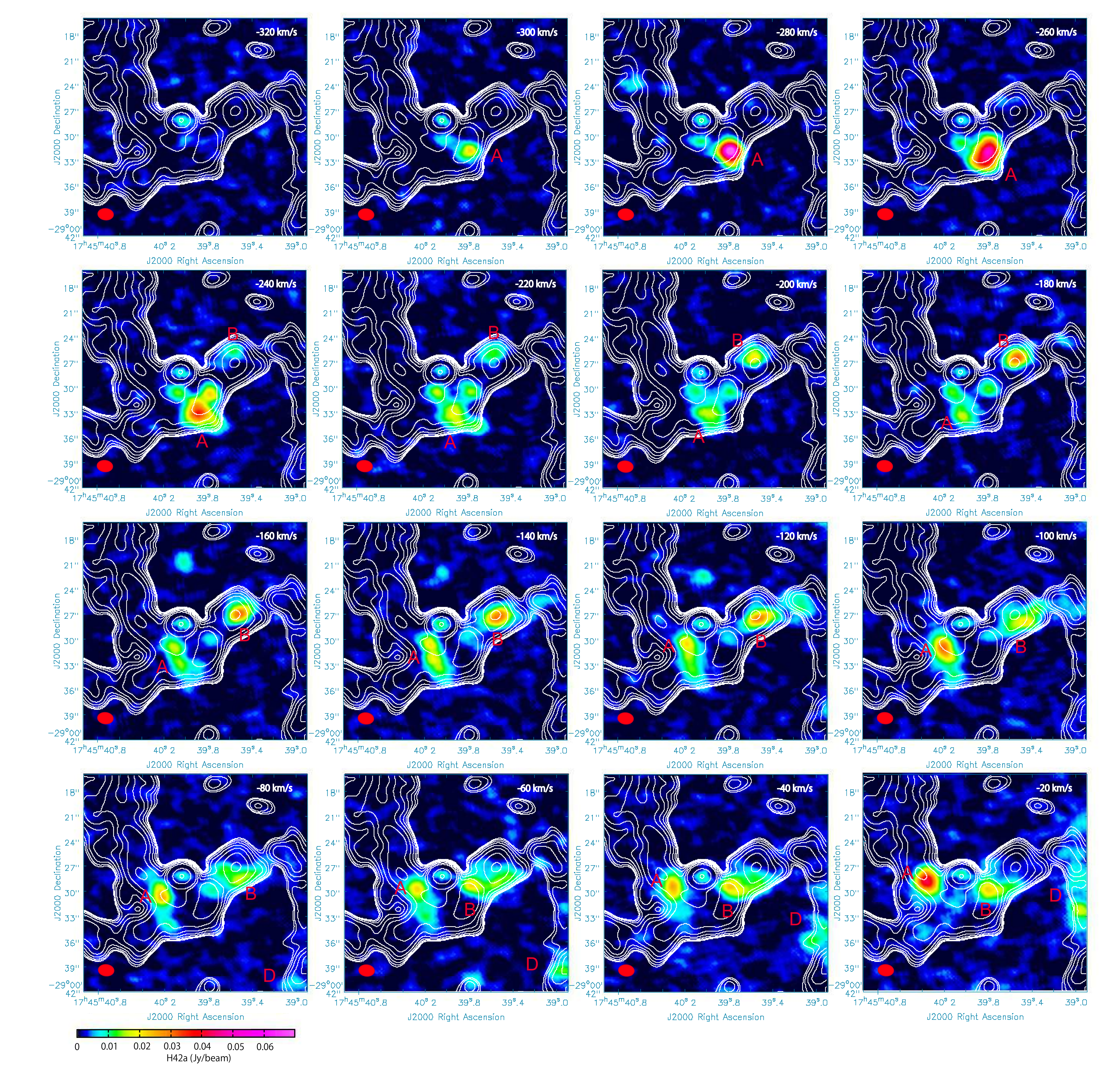}
\caption{Enlarged channel maps of the vicinity of  Sgr A$^\ast$ in the H42$\alpha$ recombination line (pseudo color). The angular resolution is $1.9\arcsec \times 1.3\arcsec$($PA=84^\circ$) using``Briggs weighting (R=0.5)", which is shown on the lower-left corner of each panel as a red filled oval.  The contours in the figure show the continuum emission of the ``Galactic Center Mini-spiral" at 100 GHz for comparison \citep{Tsuboi2016}. ``A", ``B",``C", and ``D" in the figure indicate the Northern Arm, the Bar, the Eastern Arm, and the Western Arc, respectively. }
\end{center}
\end{figure}
\begin{figure}
\figurenum{3}
\begin{center}
\includegraphics[width=18cm]{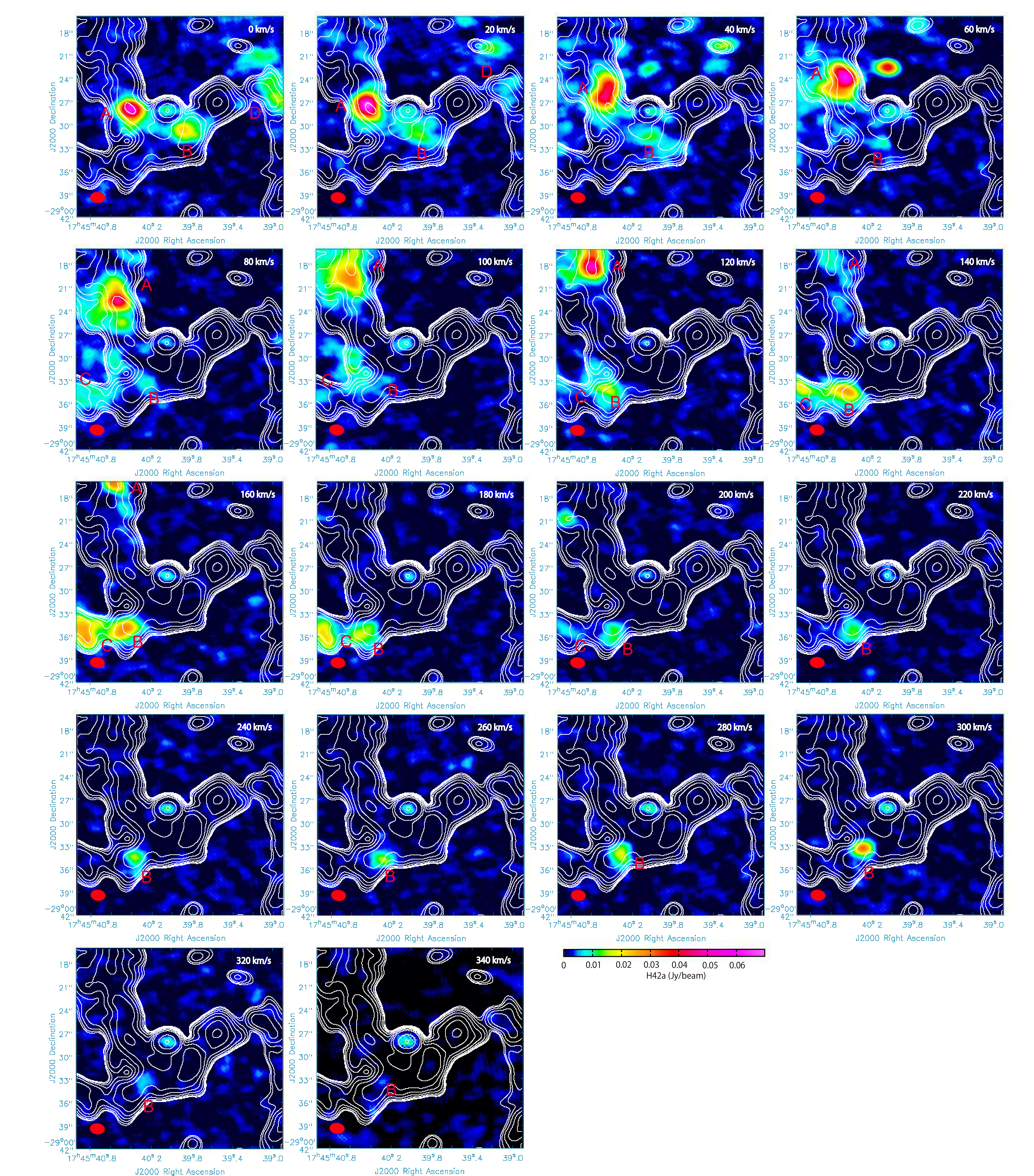}
\caption{continued}
\end{center}
\end{figure}

\section{Channel Maps and Position-velocity Diagrams}
Figure 1 shows an ALMA view of  the GCMS  at 100 GHz \citep{Tsuboi2016} for a finding chart of the substructures of the GCMS. The continuum emission in the figure is expected to be mainly originated from the ionized gas through f-f emission mechanism except for Sgr A$^\ast$ itself.  In this paper we use the traditional nomenclature of the substructures, which has been used since 1980 (features with white labels ). 
  Because the features with red labels have no customarily-called names they are newly named in the following section: Northeastern Arm and Horizontal Arm.  

Figure 2 shows the channel maps of the GCMS shown in figure 1 in the H42$\alpha$ recombination line (pseudo color). This data cube is cut from the large area mosaic data mentioned in the previous section. 
The velocity coverage of the figure is from $V_{\mathrm{LSR}}=-480$ km s$^{-1}$ to $V_{\mathrm{LSR}}=+400$ km s$^{-1}$. The velocity width of each panel is 20 kms$^{-1}$. This is comparable to the thermal velocity width of the ionized hydrogen gas, $\Delta V_{\mathrm th} = \sqrt{\frac{8ln2kT_{\mathrm e}}{m_{\mathrm H}}}\sim 21$ km s$^{-1}$ at $T_{\mathrm e}=1\times10^4$ K. The central velocity of each panel is indicated on the upper right corner.  The contours in the figure show the continuum emission of the GCMS at 100 GHz for comparison (see Figure 1). 
Although the resultant angular resolution using ``natural weighting" in figure 2, $2.5\arcsec \times 1.9\arcsec$, is not so high compared to those of previous observations;   e.g. $1.3\arcsec \times 1.3\arcsec$ for H92$\alpha$ recombination line in \cite{Zhao2009}, $1.9\arcsec \times 1.5\arcsec$ for H30$\alpha$ recombination line in \cite{Zhao2010}, the sensitivity for a 20 km s$^{-1}$ channel is as small as  $\sim0.7$ mJy beam$^{-1}$, which is improved several times than the previous values; e.g. $\sim 6$ mJy beam$^{-1}$ for H92$\alpha$ recombination line calculated from \cite{Zhao2009}, $\sim 9$ mJy beam$^{-1}$ for H30$\alpha$ recombination line calculated from \cite{Zhao2010}. 
The resolve-out scale of our maps is as large as $82\arcsec$ because of ample very short baselines by the ACA. The high sensitivity and large resolve-out scale improve the imaging fidelity.   The ionized gas components corresponding to the substructures of the GCMS are found in the velocity range from $V_{\mathrm{LSR}}=-400$ km s$^{-1}$ to $V_{\mathrm{LSR}}=+340$ km s$^{-1}$. 
In addition, there are faint components around the east edge of the panels of $V_{\mathrm{LSR}}=-200 $~to~$ -180$ km s$^{-1}$ and $\sim30\arcsec$ southeast of Sgr A$^\ast$ in the panel of $V_{\mathrm{LSR}}=-220$ km s$^{-1}$. These emissions are probably contaminations of the $^{29}$SiO $v=0~J=1-0$ emission line  ($85.759188$ GHz) because they have no continuum counterparts while the counterparts in the CS and SiO emission lines are prominent \citep{Tsuboi2016b}. On the other hand, the He42$\alpha$ recombination line ($85.7233$ GHz) is not detected in the channel maps. The non-detection shows that the number ratio of He$^+$ to H$^+$ is less than $\frac{N(\mathrm{He^+})}{N(\mathrm{H^+})} \lesssim0.1$.

Figure 3 shows the enlarged channel maps of the vicinity of  Sgr A$^\ast$ in the H42$\alpha$ recombination line (pseudo color) to clarify the kinematics of the innermost part of the ionized gas streamers. The angular resolution is $1.9\arcsec \times 1.3\arcsec, PA=84^\circ$ using ``briggs weighting (R=0.5)". The sensitivity is $\sim0.7$ mJy beam$^{-1}$*20 km s$^{-1}$. In addition, we see a faint component at the position of Sgr A$^\ast$ in Figures 2 and 3. This is probably the residual emission in the continuum subtraction process mentioned in the previous section although there is a possibility that this is the H42$\alpha$ emission surrounding Sgr A$^\ast$.
Figure 4 shows the position-velocity diagrams along the substructures of the GCMS in the H42$\alpha$ recombination line.  The integration area is shown as a rectangle in each guide map (contours).   The angular offset is measured along the long side.  The velocity bin width of each panel is 10 km s$^{-1}$. 

\begin{figure}
\figurenum{4}
\begin{center}
\includegraphics[width=18cm]{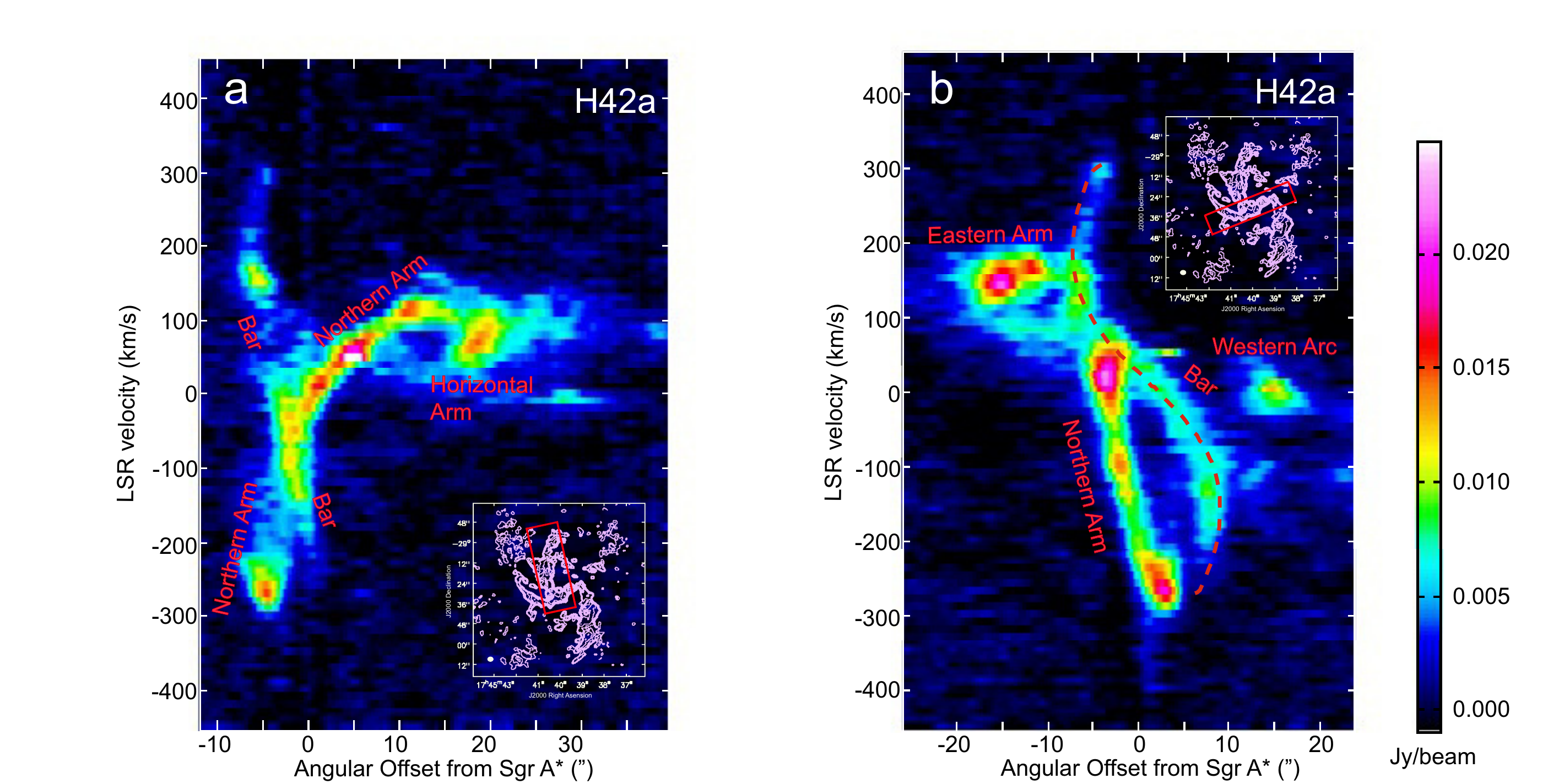}
\caption{Position-velocity diagrams along the substructures of the ``Galactic Center Mini-spiral (GCMS)" in the H42$\alpha$ recombination line. {\bf a} Position-velocity diagram along the Northern Arm. The integration area is shown as a rectangle in the guide map  (contours). {\bf b} Position-velocity diagram along the Bar. The integration area is shown as a rectangle in the guide map  (contours). 
  Broken line shows a fitted curve with $e=0.8$ at $PA=0^\circ$ (see also figure 12). 
 }
\end{center}
\end{figure}

\subsection{the Northern Arm}
The Northern Arm (NA) is the most prominent continuum substructure of the GCMS (see Figure 1). The ionized gas streamer corresponding to the NA is detected in the channel maps with the velocity range  from $V_{\mathrm{LSR}}\sim-300$ km s$^{-1}$ to $V_{\mathrm{LSR}}\sim+160$ km s$^{-1}$(see Figures 2 and 4a).  
The most negative velocity component of the NA is seen in the channel maps with the velocity range of $V_{\mathrm{LSR}}\sim-300 $~to~$ -220$ km s$^{-1}$(see ``A" in Figure 3 and a local peak in Figure 4a). This component is located around $\alpha\sim17^h45^m39.8^s, \delta\sim-29^\circ00\arcmin32\arcsec$ and appears to be associated with IRS 2L. The component has been identified in previous observations \citep[e.g.][]{Zhao2009, Zhao2010}.  In the velocity range of $V_{\mathrm{LSR}}\sim-300 $~to~$ -180$ km s$^{-1}$, the intensity peak position of the component shifts to southeast according as the velocity goes to positive. Then the peak  position with $V_{\mathrm{LSR}}\sim-180$~to~$+160$ km s$^{-1}$ shifts to north along the NA with increasing velocity. 
The ionized gas streamer is clearly identified as a curved ridge  in the position-velocity diagrams along the NA (Figure 4a), which is connecting the north end with $V_{\mathrm{LSR}}\sim+100$ km s$^{-1}$ and the southwest end with $V_{\mathrm{LSR}}\sim-300$ km s$^{-1}$. The ionized gas streamer is also identified as  a slightly curved ridge with a large velocity gradient  in the position-velocity diagram along the Bar (Figure 4b).
In addition, an extended component with the velocity range from $V_{\mathrm{LSR}}\sim+40$ km s$^{-1}$ to $V_{\mathrm{LSR}}\sim+140$ km s$^{-1}$ is also seen in the whole of the NA (see figure 2). 

There are some components apparently connecting with the NA in the channel maps.
A curved component  appears abruptly around the north end of the NA  in the panels of $V_{\mathrm{LSR}}\sim-20$~to~$0$ km s$^{-1}$ of Figure 2. The ionized gas component probably corresponds to IRS 8. This component is seen as a narrow velocity width component with $V_{\mathrm{LSR}}\sim-20$~to~$ 0$ km s$^{-1}$ at $20\arcsec $~to~$ 30\arcsec$ angular offsets in the position-velocity diagram along the NA (see Figure 4a). The component should not be physically associated with the ridge of the NA because it is isolated from the NA in the position-velocity diagram. 
Another ionized gas streamer crosses the NA around $\alpha\sim17^h45^m40.7^s, \delta\sim-29^\circ00\arcmin11\arcsec$ at nearly right angle and is called ``Horizontal Arm (HA)" hereafter. The peak position of this streamer moves to east according as the velocity increases  from $V_{\mathrm{LSR}}\sim -40$ km s$^{-1}$ to $V_{\mathrm{LSR}}\sim+100$ km s$^{-1}$. The HA is also identified as a component with a velocity width of $\Delta V\gtrsim100$ km s$^{-1}$ at $\sim20\arcsec$ angular offset in Figure 4a. The component has been identified in previous observations \citep[e.g.][]{Zhao2009, Zhao2010}. 

\subsection{the Bar} 
The ionized gas streamers seen toward the Bar have two distinct velocity structures, which are shown in the position-velocity diagram along the Bar (see Figure 4b).
The first one has an elongated S-shaped structure (curved broken line) in the position-velocity diagram, which has large velocity width features at the both velocity ends. This structure is also seen as an inclined linear feature in the position-velocity diagram along the NA (see Figure 4a). 
The second one has a curved ridge with a large velocity gradient, which is prominent around the angular offset of  $\sim0\arcsec$ from Sgr A$^\ast$ in the velocity range of $V_{\mathrm{LSR}}\sim-300 $~to~$ 60$ km s$^{-1}$.  This component is identified as the NA mentioned in the previous subsection. 
Then the first component is the ionized gas streamer corresponding to the Bar itself. This component is also identified in the channel maps (``B" in Figure 3). 
The negative velocity end of the component appears at $\alpha\sim17^h45^m39.6^s,\delta\sim-29^\circ00\arcmin26\arcsec$ in the panel of  $V_{\mathrm{LSR}}=-240$ km s$^{-1}$. 
The intensity peak  of the component stays around the continuum peak, $\alpha\sim17^h45^m39.6^s,\delta\sim-29^\circ00\arcmin27\arcsec$ (see Figure 1), although it slightly shifts southwest according as the velocity goes to positive in the panels of $V_{\mathrm{LSR}}=-240 $~to~$ -100$ km s$^{-1}$. Next the peak shifts to southeast along the Bar according as the velocity increases from $V_{\mathrm{LSR}}=-80$ km s$^{-1}$ to $V_{\mathrm{LSR}}=+120$ km s$^{-1}$. The component  stays around the same position in the velocity range from $V_{\mathrm{LSR}} = +120 $~to~$ +220$ km s$^{-1}$ and may not mingle with the EA. In the panels of $V_{\mathrm{LSR}}=+220 $~to~$ +340$ km s$^{-1}$, a compact component is seen around $\alpha\sim17^h45^m40.3^s,\delta\sim-29^\circ00\arcmin34\arcsec$, i.e., at the southeast edge of the Mini-cavity.
This is presumably the positive velocity end of the Bar.  The kinematics of the Bar is discussed in section 4.

\begin{figure}
\figurenum{4}
\begin{center}
\includegraphics[width=18cm]{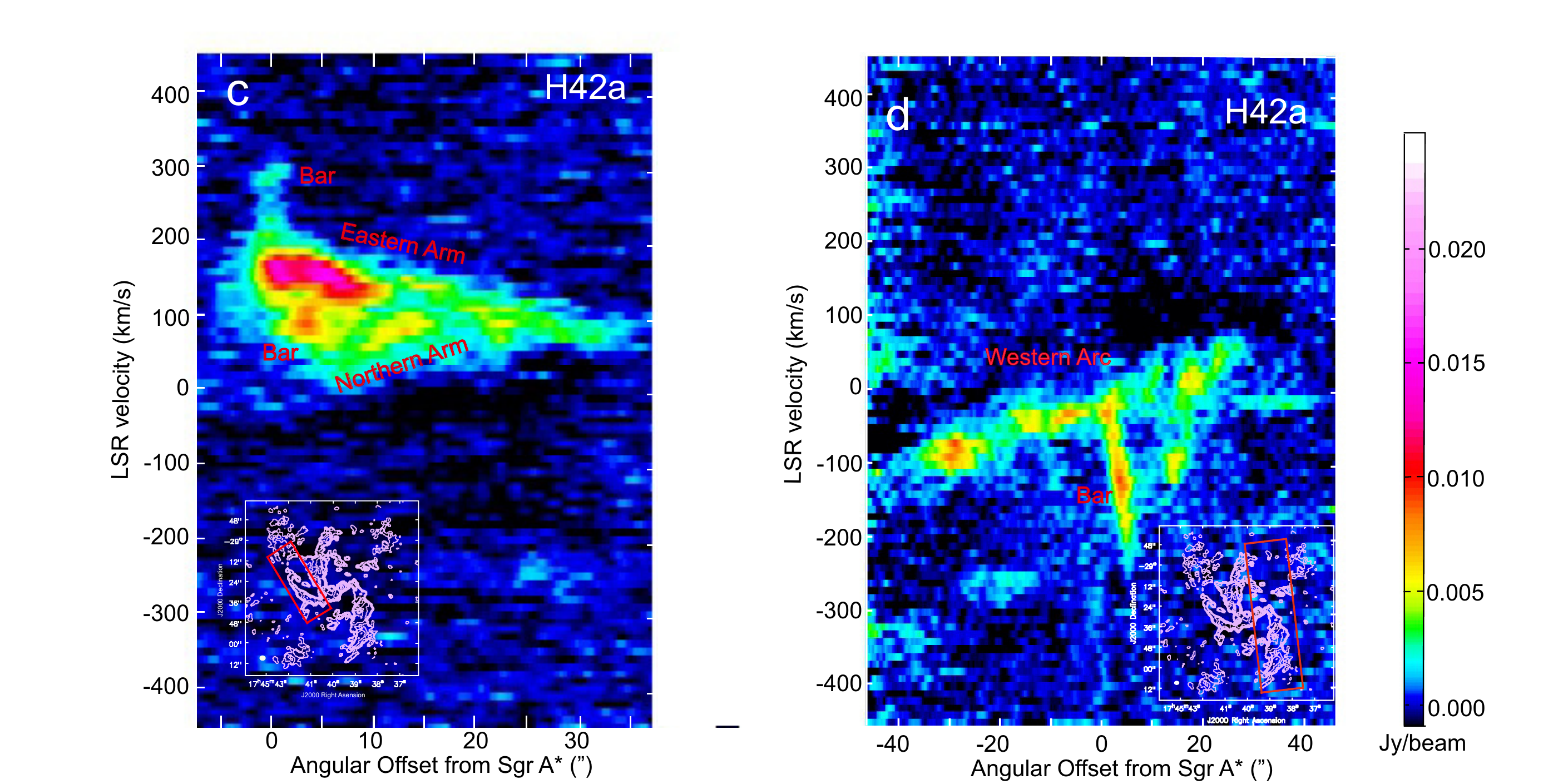}
\caption{(continued.) {\bf c} Position-velocity diagram along the Eastern Arm in the H42$\alpha$ recombination line. The integration area is shown as a rectangle in the guide map  (contours). {\bf d} Position-velocity diagram along the Western Arc  in the H42$\alpha$ recombination line. The integration area is shown as a rectangle in the guide map  (contours). 
}
\end{center}
\end{figure}

\subsection{the Eastern Arm}
The EA is also a conspicuous continuum substructure of the GCMS (see Figure 1). 
The ionized gas streamer corresponding to the EA has complicated velocity structures.
The negative velocity end of the streamer appears as an elongated component apparently connecting to the Bar in the panel of $V_{\mathrm{LSR}}=80$ km s$^{-1}$ in Figure 2. 
The component extends along the EA in the panels of $80-140$ km s$^{-1}$ and reaches at least up to $\alpha\sim17^h45^m39.5^s,\delta\sim-28^\circ59\arcmin36\arcsec$. The peak of the component seems to shift to west with going to positive velocity in the panels of $160 $~to~$ 200$ km s$^{-1}$.
The velocity structure is seen as an inclined linear ridge in the position-velocity diagram along the EA (see Figure 4c), of which velocity increases from $V_{\mathrm{LSR}}\sim80$ km s$^{-1}$at the north end to $V_{\mathrm{LSR}}\sim200$ km s$^{-1}$at the south end. The velocity width increases around the south end. 
We also see the feature in the position-velocity diagram along the Bar (see Figure 4b). The EA apparently overlaps with the Bar in the diagrams, but this is presumably coincidence in the line-of sight by chance.

Another ionized gas streamer is seen $7\arcsec$ north of the EA (see Figure 1) and is called ``Northeastern Arm (NEA)" hereafter. The NEA is also identified in the panels of $V_{\mathrm{LSR}}=60$~to~$160$ km s$^{-1}$  (see Figure 2). 
The  NEA appears to cross the EA at its north end around $\alpha\sim17^h45^m41.9^s,\delta\sim-29^\circ00\arcmin20.5\arcsec$  (see Figure 1). The negative velocity end of the  NEA appears as a curved extension to $\sim10\arcsec$ northeast of the crossing point in the panel of $V_{\mathrm{LSR}}=60$ km s$^{-1}$. 
 The velocity component tends to shift to southwest along the NEA with increasing velocity. The positive velocity end of the ionized gas is seen as an elongated source between the NA and the EA in the panel of $V_{\mathrm{LSR}}=160$ km s$^{-1}$. However, the inner end of the NEA does not appear to reach to the vicinity of Sgr A$^\ast$. The southwest end corresponds to the group of half-shell like sources found by JVLA at 34 GHz \citep{Yusef-Zadeh2015}, which is discussed in section 4.

\subsection{the Western Arc}
 
Figure 2 shows that the intensity peak goes to north along the Western Arc (WA) with increasing velocity from $V_{\mathrm{LSR}}=-140$ to $40$ km s$^{-1}$.  The component of the WA with $V_{\mathrm{LSR}}=-20$ to $20$ km s$^{-1}$ appears to cross the Bar.
Figure 4d shows the position-velocity diagram along the WA. The WA is seen as an inclined linear feature crosses the point of $0$ km s$^{-1}$ and $0\arcsec$ offset in the diagram. This linear feature suggests that the WA is a part of a nearly circular orbiting ring \citep[e.g.][]{Zhao2009}. 
 The Bar is also seen as a linear feature with a high velocity gradient at the angular offset of $\sim0\arcsec$ from Sgr A$^{\ast}$ in the diagram.  Because both components  cross at nearly right angle in the diagram, the WA would not be physically associated with the Bar.

\begin{figure}
\figurenum{5}
\begin{center}
\includegraphics[width=15cm]{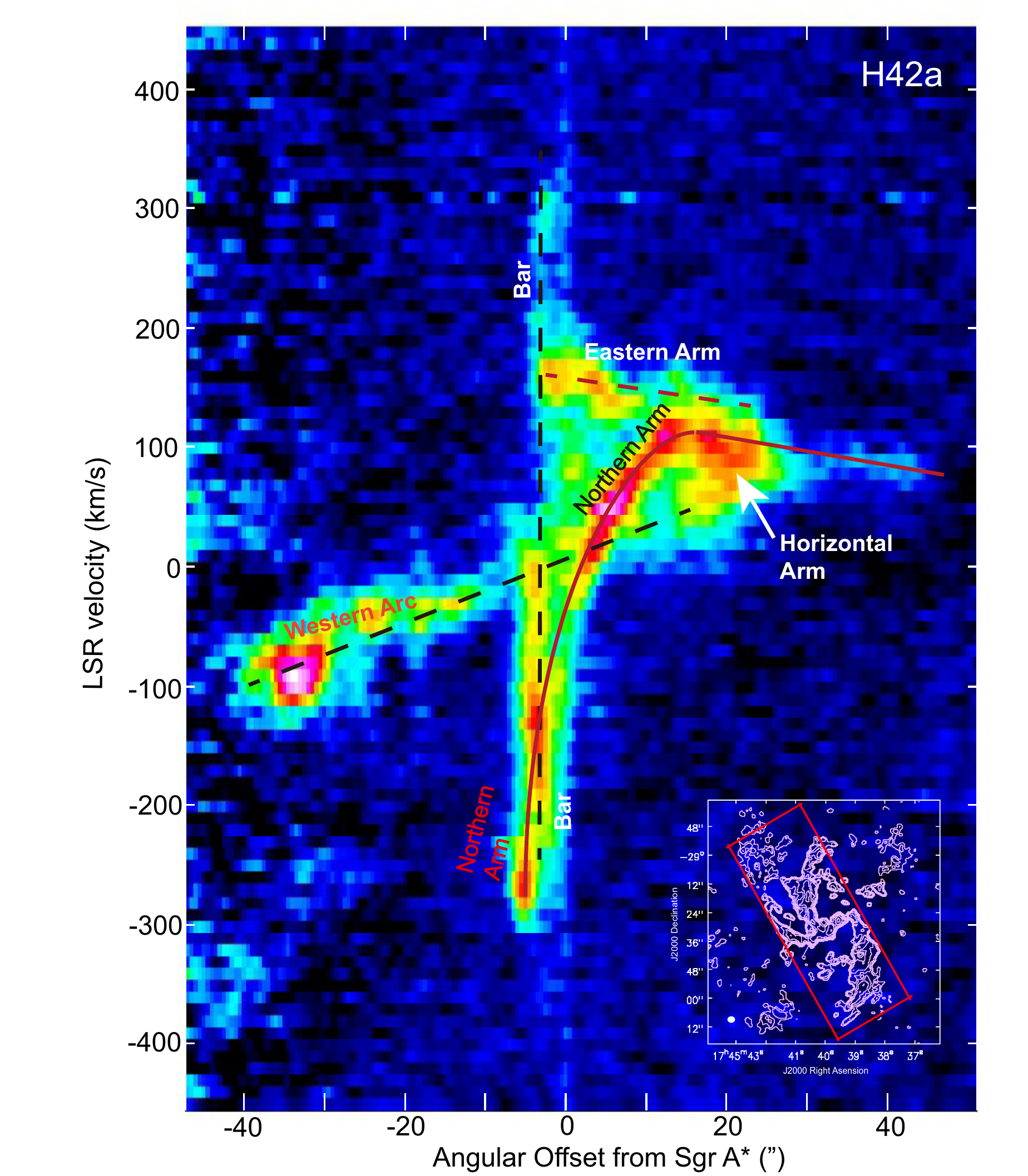}
\caption{ The finding chart of the ionized gas streamers on the position-velocity diagram along galactic longitude in the H42$\alpha$ recombination line. 
}
\end{center}
\end{figure}
\begin{figure}
\figurenum{6}
\begin{center}
\includegraphics[width=16cm]{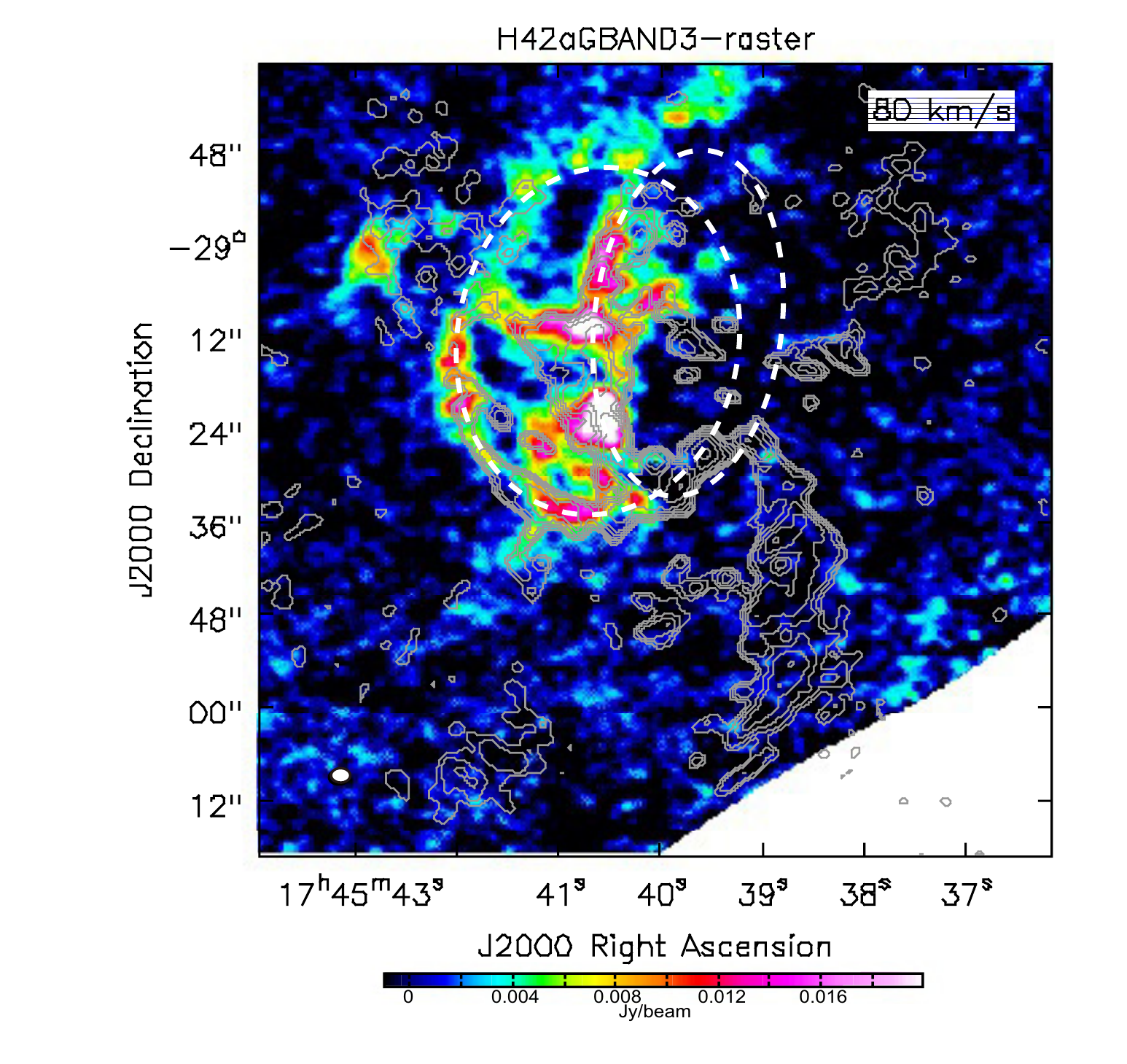}
\caption{Comparison between  the H42$\alpha$ recombination line intensity of the velocity range from $V_{\mathrm{LSR}}=70$ to $90$ km s$^{-1}$  (pseudo color) and the trajectories of the elliptical Keplerian orbits for the Norther Arm and the Eastern Arm and Bar, which were proposed by \cite{Zhao2009} (white broken line ovals).}
\end{center}
\end{figure}
\section{Discussion}
\subsection{Kinematics of the Northern Arm and  the Eastern Arm}
Figure 5 shows the position-velocity diagram along galactic longitude in the H42$\alpha$ recombination line.
This  is the finding chart of the ionized gas streamers discussed in the following. The diagram shows the mutual relations among their velocity structures of the gas streamers.

A model that the ionized gas streamers of the GCMS are in elliptical Keplerian orbits has been advocated to explain the observed kinematics and structures of the GCMS \citep[e.g.][]{Zhao2009}.  
The ionized gas  should not be bounded by self gravity in the GCMS as well as in the HII regions of the Galactic disk. 
When the ionized gas in the GCMS moves along its Keplerian orbit, the ionized gas can expands up to $D\sim1$ pc within one orbital period, which is comparable to the semi-major axis of the orbit of $\sim1$ pc, because the typical thermal velocity and  typical orbital period are $\sim10$ km s$^{-1}$ and $\sim10^4$ yr, respectively. This corresponds to the angular diameter of $D\sim20\arcsec$. 
However, the observed morphology of the ionized gas appears the trajectory of the orbit at least within the scope of this observation. Although the width of the ionized gas is wider than that of the dust ridge in the GCMS \citep{Tsuboi2016b}, there would be no sign of such rapid expansion on the ionized gas streamers (see Figure 1 and Figure 2). Some external pressure is necessary in order to prevent the expansion of the ionized gas. 
The electron density of ambient ionized gas can be estimated from the observations by single-dish radio telescopes which detected the extended emission \citep[e.g][]{Mezger1986,Tsuboi1988}. 
The electron temperature is assumed to be $T_e=1\times 10^4 $ K here because it has been reported to be $T_e\sim5-13\times 10^3 $ K in the region \citep{Mezger1986}.  The ambiguity of the electron temperature  makes the error of at most $\pm$10 \% in the electron density because the electron density has a small dependence to the electron temperature (see equation 2). The averaged ambient electron density within $0.5$ pc is derived to be $n_e\sim2\times10^3$ cm$^{-3}$ from the continuum flux density at 91 GHz subtracting the GCMS and Sgr A$^\ast$  \citep[$S_{\nu, \mathrm{amb}}=3.8$ Jy beam$^{-1}$($20"$), see equation 1, ][]{Tsuboi1988}. 
The electron density of the NA by our ALMA observation, $n_e\sim7-13\times10^3$ cm$^{-3}$ (see Table 1),  is at most several times denser than the ambient electron density.  The small contrast between these densities suggests that the ambient ionized gas plays some role in the confinement of the ionized gas streamers. Moreover, the ambient electron density should increase in proportion to $r^{-0.5}$  in the case of Bondi accretion, here $r$ is the distance from Sgr A$^\ast$. The degree of the confinement will increase with approaching the periastron.

On the other hand, such ambient ionized gas may also affect somewhat the Keplerian orbital motion by Ram pressure because  the orbital velocity is as large as $v\sim100$ km s$^{-1}$.  Here we estimate the ram pressure of a gas globe with  the radius of $R\sim0.1$ pc at the distance of $r\sim0.5$ pc from Sgr A$^\ast$, and compare this with the gravity by Sgr A$^\ast$. The perturbation by ram pressure is estimated to be $f_{\mathrm{ram}}=n_{\mathrm{e}}{\mathrm{(ambient)}}m_{\mathrm{p}}v^2\times\pi R^2\sim1\times10^{29}$ g cm s$^{-2}$.
On the other hand, the gravity by Sgr A$^\ast$ is estimated to be $f_{\mathrm{g}}=\frac{GM}{r^2} n_{\mathrm{e}}{\mathrm{(GCMS)}}m_{\mathrm{p}} \times \frac{4}{3}\pi R^3 \sim4\times10^{30}$ g cm s$^{-2}$ assuming that  the density is $n_e{\mathrm{(GCMS)}}\sim1\times10^4$ cm$^{-3}$. The deceleration effect on the Keplerian orbital velocity is a few \% of the gravity at the distance.
If this is the case, the deceleration effect would increase with approaching the periastron, and the semi-major axis by fitting to the inner part of the trajectory would be smaller than that by fitting to the outer part. 

Figure 6 shows the comparison between the ionized gas distribution of $V_{\mathrm{LSR}}=70-90$ km s$^{-1}$ and the trajectories of the elliptical Keplerian orbits for the NA and EA+Bar which were proposed by Zhao et al (2009).  
The correspondence between the ionized gas distribution and the proposed trajectory is very well in the EA. The northern half  trajectory of the EA found  by our observation also shows the clear correspondence.
Although the EA is seen to connect with the Bar in the continuum emission map and the integrated intensity map of the H42$\alpha$ recombination line (see Figure 1 and Figure 10), the velocity structures of the EA and the Bar do not smoothly connect with each other in the PV diagram (see Figure 5). Then the ionized gas distribution in the Bar does not fit the trajectory well. The kinematics of the Bar is discussed latter. 
Meanwhile, although the ionized gas distribution fits the proposed trajectory in the inner part of the NA,  the distribution appears to extend to north beyond the  proposed trajectory for the NA. The extension would be understood as  a northernmost part of the NA in the original orbit that was not perturbed by the ambient gas.

An alternate model that the NA and WA form a one-armed spiral structure in a nearly circular orbit also has been proposed, which is based on the observations in the [NeII] emission line \citep[e.g.][] {Lacy1991, Irons}.  In this model, the ionized gas of the NA and WA  does not flow along the features but flows at a large angle to them. In addition, the NA and WA should connect with each other around $\alpha=17^h45^m40^s, \delta=-29^ \circ00\arcmin 08\arcsec$ \citep[see Fig.1 in][] {Irons}.  
However, the NA appears to extend north beyond the connecting point in the  H42$\alpha$ recombination line (see Figure 5). Note that the northern extension of the NA is identified even in the integrated intensity map in the [NeII] emission line \citep[see Fig.1 in][] {Lacy1991}.  Moreover, the velocity structures of the NA and the WA do not smoothly connect with each other in the PV diagram. The velocity structure of the HA apparently bridges the gap between the WA and the halfway of the NA (see also Figure 4a). 

\begin{figure}
\figurenum{7}
\begin{center}
\includegraphics[width=16cm]{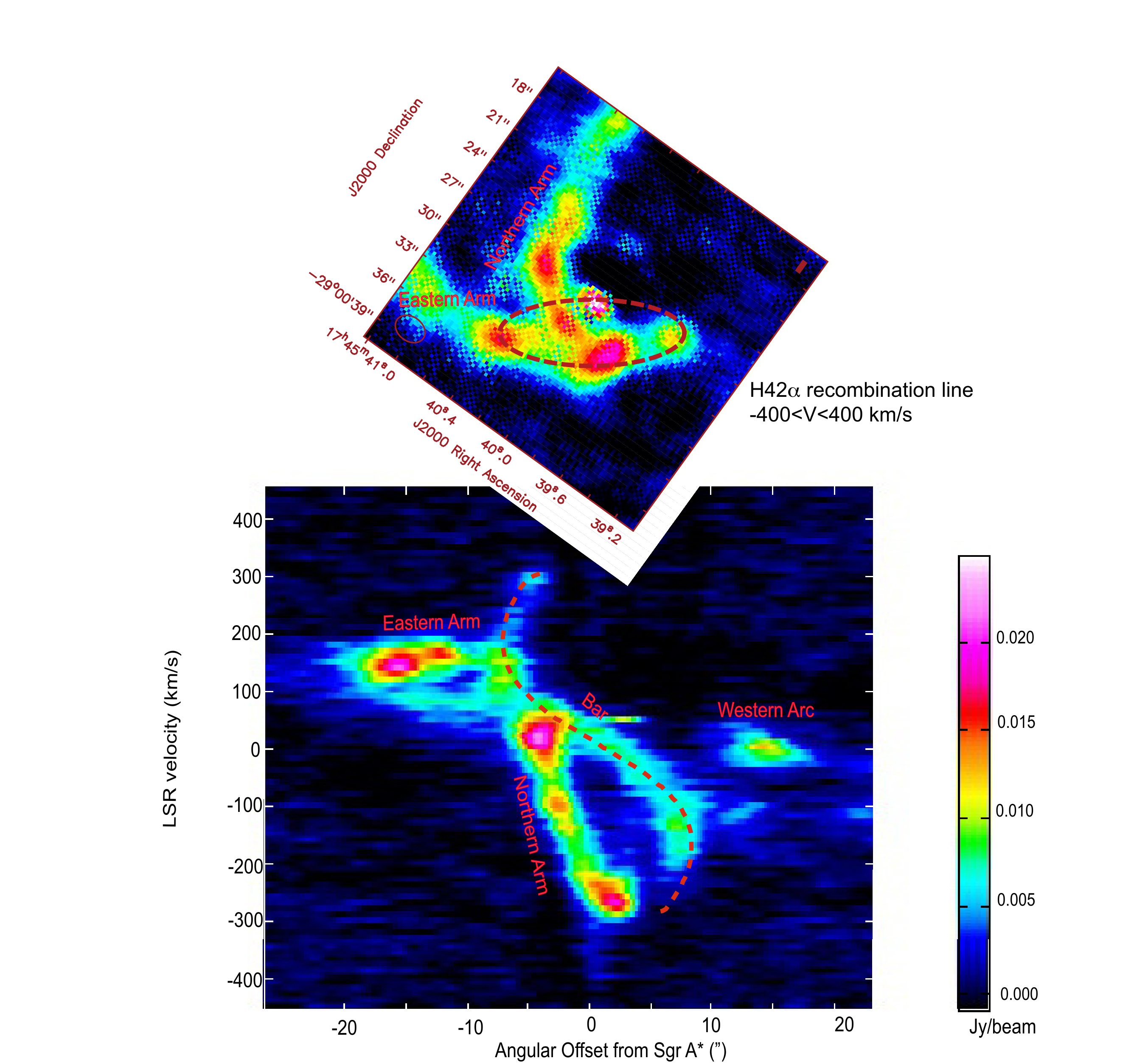}
\caption{{\bf (upper panel)}  the integrated intensity map of the H42$\alpha$ recombination line. The integrated velocity range is $V_{\mathrm{LSR}}=-400$ to $+400$ km s$^{-1}$. The broken line shows the Keplerian orbit with $e= 0.8$ at $PA=0^\circ$. {\bf (lower panel)}  the position-velocity diagram along the Bar. The streamer belong to the Bar is  appeared as a S-shaped ridge.  on the position-velocity diagram along the Bar (see also Figure 4b). The kinematic characters are reproduced only by the Keplerian orbit with $e= 0.8$ and $i= 80^\circ$ at $PA=0^\circ$. }
\end{center}
\end{figure}

\subsection{Kinematics of the Bar }
 Two distinct models have been proposed to explain the kinematics and structures of the Bar. One model is that the Bar is understood as components in the elliptical Keplerian orbit including the EA \citep[e.g.][]{Zhao2009}. However, some discrepancies between the model and the observed kinematics have been indicated \citep[e.g.][]{Irons}.  
Another model is that the Bar is understood as components in a Keplerian orbit not including the other streamers, which is nearly edge-on \citep{Liszt}. The model is based on the observation of the radio recombination line, H92$\alpha$ \citep{Roberts}, which covered the  $V_{\mathrm{LSR}}$ range of $-200$ to $+200$ km s$^{-1}$, narrower than that in our observation. 

Our observation supports the latter model.
In figure 3, the component B shifts along the minor axis of the elongated Bar structure with increasing velocityfrom $V_{\mathrm{LSR}}=-240$  to $-80$ km s$^{-1}$. Such positional shift cannot be understood as a part of the previously proposed Keplerian orbit for the EA+Bar.  On the other hand this positional shift can be explained by an independent Keplerian orbit with high eccentricity based on our observation.  The component B approaches Sgr A$^{\ast}$ along the major axis of the elongated Bar structure with increasing velocity from $V_{\mathrm{LSR}}=200$  to $340$ km s$^{-1}$. 

Figure 7 shows the relation between the S-shaped ridge appeared on the position-velocity diagram along the Bar (see also Figure 4b) and the Keplerian trajectory in the Bar, which is derived in the following part.
The upper panel is the integrated intensity map of the H42$\alpha$ recombination line and the lower panel is  the position-velocity diagram along the Bar.
The kinematic characteristics on the position-velocity diagram suggest that the S-shaped ridge is a part of a component with a nearly Keplerian orbit around the GCBH, of which the major axis should be nearly parallel to the line-of-sight. Position-velocity diagrams of Keplerian orbits with high eccentricity are calculated in APPENDIX. As shown in figure 12, the kinematic characteristics seen in the figure are reproduced only in the case of $PA\sim0^\circ$.    
Because the ionized gas streamer cannot be confined simply to a single orbit by the perturbation mentioned in the previous subsection, and the gas distribution is not homogeneous, it is not easy to derive the orbital parameters accurately from the characteristics of the streamer on the position-velocity diagram. 
However, the orbit probably has a high eccentricity. The curves with $e= 0.8$ seems to well reproduce the observed position-velocity curve of the Bar. The broken line in the lower panel shows the orbit with $e= 0.8$ and $i= 80^\circ$(please see the following paragraph) at $PA=0^\circ$. 

The semi-minor axis of the orbit is expected to be $b=a\sqrt{1-e^2}\sim0.3$ pc from the angular extent on the position-velocity diagram, $\theta\sim\pm8\arcsec$. The semi-major axis and period of the orbit are  $a\sim0.5$ pc and $P\sim1.7\times 10^4$ years, respectively.   Assuming the  GCBH mass  of  $M_{\mathrm {GC}}\sim4\times 10^6$ M$_\sun$, the maximum radial velocity is estimated to be $V_{\mathrm {max}}=\sqrt{\frac{GM}{a(1-e^2)}}=304$ km s$^{-1}$. The radial velocities at the maximum angular offsets are estimated to be $\mp\sqrt{\frac{GM}{a}}\sim\mp180$ km s$^{-1}$. These are actually seen on the position-velocity diagram. Comparing the calculated maximum velocity with the maximum velocity observed on the position-velocity diagram, the angle between the line-of-sight and the normal of the orbital plane,  that is the inclination angle, is as large as $i\sim80^\circ$. The semi-major axis and the inclination angle estimate that the projected angular extent along the semi-major axis is $\lesssim5\arcsec$. The expected orbit with $i=80^\circ$  is shown as an oval on the integrated intensity map (see the upper panel of figure 7).

 The component B moves along the orbit in the velocity range from $V_{\mathrm{LSR}}=-80$  to $80$ km s$^{-1}$(see figure 3). The component should be located at the opposite side of Sgr A$^\ast$ on the orbit, i.e., around the apoastron. 
The periastron distance from the GCBH and the orbiting velocity at the periastron are estimated to be $a(1-e) \sim0.1$ pc and $\sqrt{\frac{GM}{a}}\sqrt{\frac{1+e}{1-e}}\sim550$ km s$^{-1}$, respectively. The periastron is probably located within the Bondi accretion radius of Sgr A$^{\ast}$ , $R_{\mathrm A}\sim0.2$ pc, which is derived from X-ray observations by CHANDRA \citep{Wang}. 

\begin{figure}
\figurenum{8}
\begin{center}
\includegraphics[width=16cm]{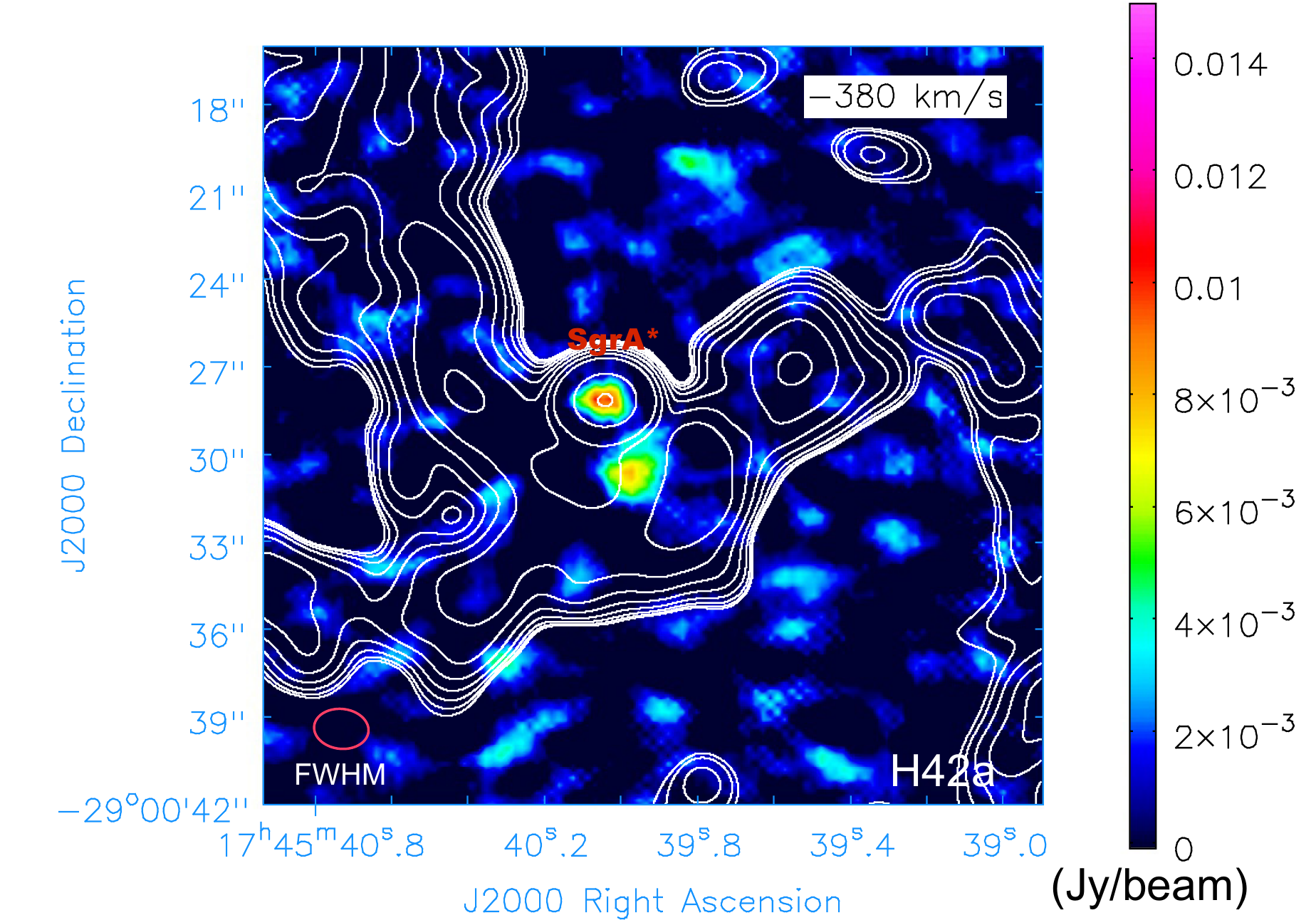}
\caption{Magnified map of the vicinity of  Sgr A$^\ast$ in the H42$\alpha$ recombination line (pseudo color). The velocity range is from $V_{\mathrm{LSR}}=-390$ to $-370$ km s$^{-1}$. The angular resolution is $1.9\arcsec \times 1.3\arcsec$($PA=84^\circ$) which is shown on the lower-left corner as an open oval.  Notes that a faint component seen at the position of Sgr A* is probably a residual of continuum subtraction by CASA.  
The contours in the figure show the continuum emission of the ``Galactic Center Mini-spiral (GCMS)" at 100 GHz for comparison \citep{Tsuboi2016}.}
\end{center}
\end{figure}
\subsection{High Velocity Component of the Galactic Center Mini-Spiral }
  Figure 8 shows a magnified map of the vicinity of  Sgr A$^\ast$ in the H42$\alpha$ recombination line. The velocity range is from $V_{\mathrm{LSR}}=-390$ to $-370$ km s$^{-1}$. A faint component with extremely negative velocity is identified at $\alpha\sim17^h45^m40.0^s, \delta\sim-29^\circ00\arcmin30.8\arcsec$,  which corresponds to $\sim2.8\arcsec$ ($\sim0.11$pc) south-southwest of  Sgr A$^\ast$.
The IR counterpart of the faint component has been identified in SINFONI observations \citep{Steiner}. Note that another faint component seen at the position of Sgr A* is probably a residual in the continuum subtraction process by CASA. 

Assuming a circular Keplerian orbit around Sgr A$^\ast$, the orbital velocity is estimated to be $V_{\mathrm orbit}=\sqrt{\frac{GM}{r}}\sin i$.
Because the observed radial velocity is consistent with the  orbital velocity estimated for a circular Keplerian orbit  with $i\sim 90^\circ$, $V_{\mathrm orbit}\sim395$ km s$^{-1}$, the physical distance from the GCBH of the component should be close to the projected distance.
The entire orbit is located within the Bondi accretion radius.
On the other hand the observed  radial velocity is also consistent with a free fall velocity ($\sim550 \sin i$ km s$^{-1}$) from $r=3$ pc, i.e., the outer end of the GCMS, to the observed position with $i\sim 45^\circ$. This component might be an inner tip of a streamer approaching to the GCBH.
This will be instantly disrupted by strong tidal shear of the GCBH and a part of the fragments will begin free-fall to it \citep[e.g.][]{Saitoh}.  

\begin{figure}
\figurenum{9}
\begin{center}
\includegraphics[width=19cm]{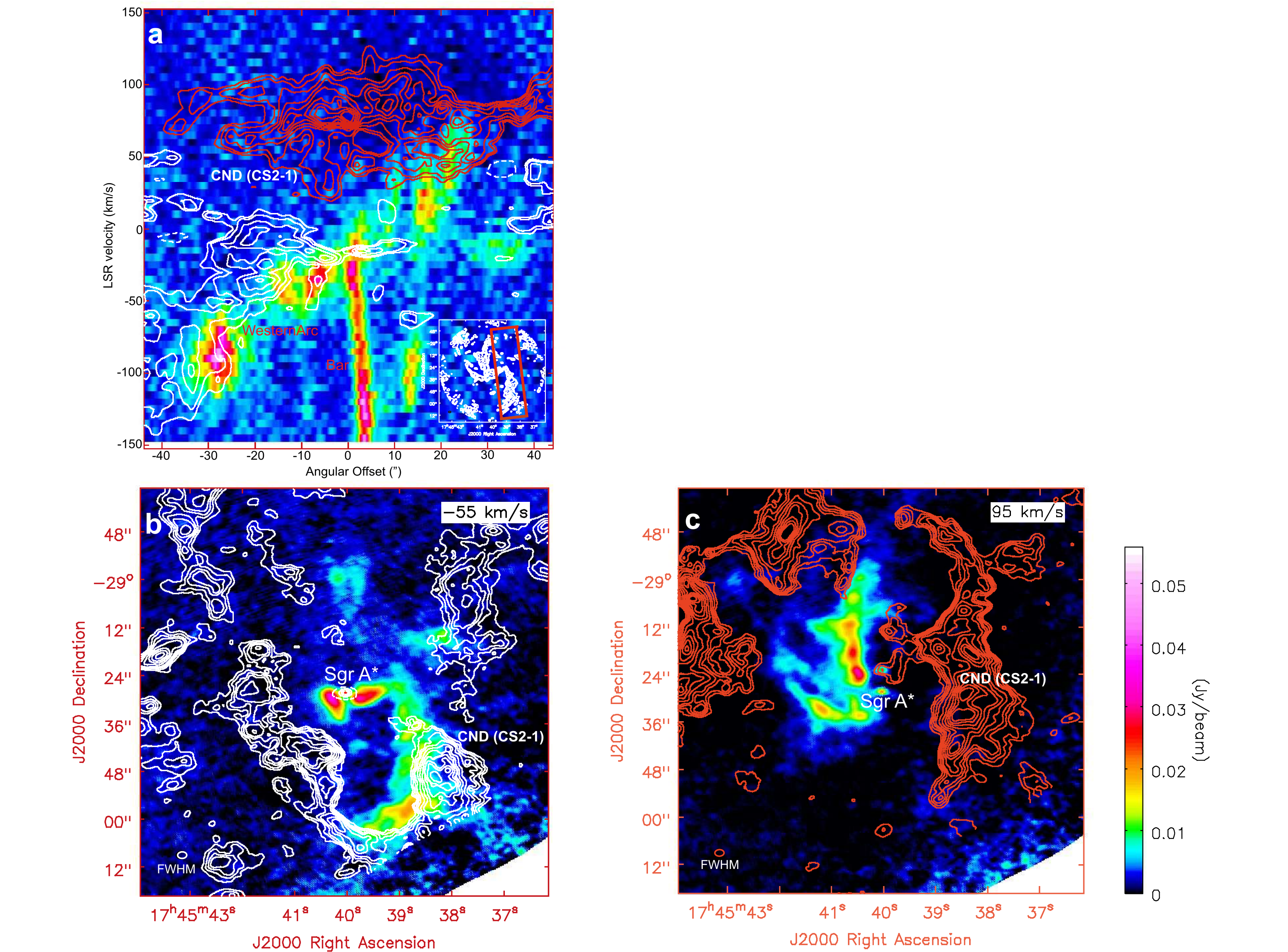}
\caption{{\bf a}  The position-velocity diagram of the Circumnucler Disk (CND) in the CS $J=2-1$ emission line \citep[contours, ][]{Tsuboi2016b} overlaied on the position-velocity diagram along the Western Arc in the H42$\alpha$ recombination line (pseudo color). {\bf b} The Circumnucler Disk (CND) in the CS $J=2-1$ emission line \citep[white contours, ][]{Tsuboi2016b} overlaied on the Western Arc in the H42$\alpha$ recombination line (pseudo color). The velocity range is $V_{\mathrm{LSR}}=-105$ to $-5$ km s$^{-1}$. {\bf c} The Circumnucler Disk (CND) in the CS $J=2-1$ emission line \citep[red contours, ][]{Tsuboi2016b} overlaied on the Northern Arm and Eastern Arm  in the H42$\alpha$ recombination line (pseudo color). The velocity range is $V_{\mathrm{LSR}}=45$ to $145$ km s$^{-1}$.}
\end{center}
\end{figure}
 \subsection{Relation between the Western Arc and the Circum-Nuclear Disk}
Figure 9a  shows the comparison between the position-velocity diagrams of the WA in the H42$\alpha$ recombination line and the CND in the CS $J-2-1$ emission line \citep[contours,][]{Tsuboi2016b}.  
There is a molecular gas component with  $V_{\mathrm{LSR}}=-100$  to $0$ km s$^{-1}$ of the Circumnuclear Disk (CND) which is along the outer boundary of the WA  (see Figure 9b)  \citep[e.g.][]{Christopher, Montero, Martin, Lau}. 
The  molecular gas component with the negative angular offset  and negative velocity is located on the just positive velocity side of the ionized gas component of the WA in the position-velocity diagram. 

Figure 9b shows the relation between the WA in the H42$\alpha$ recombination line  (pseudo color) and the CND in the CS $J-2-1$ emission line \citep[white contours, ][]{Tsuboi2016b} in the velocity range of $V_{\mathrm{LSR}}=-105$ to $-5$ km s$^{-1}$.  
This molecular gas component is located just outside the western periphery of the WA \citep[e.g. ][]{Christopher, Montero, Martin}.  
Thus, it is plausible that the ionized gas  component is physically associated with the molecular gas component with negative velocity. 
Moreover, the overall motion of the ionized and molecular gas is nearly circular rotation around Sgr A$^\ast$  with a velocity of $\sim100$ km s$^{-1}$.
 
Meanwhile there is another molecular gas component of the CND  in the velocity range from $V_{\mathrm{LSR}}=60$  to $110$ km s$^{-1}$ (see Figure 9a).  Figure 9c shows the molecular gas component of the CND \citep[red contours, ][]{Tsuboi2016b} overlaied on the H42$\alpha$ recombination line (pseudo color) with the velocity range of $V_{\mathrm{LSR}}=45$ to $145$ km s$^{-1}$.
The component has no ionized gas counterpart with the same velocity although it is located just outside along the western periphery of the WA traced by the mm-wave continuum emission (see also Figure1). 
It has long been advocated that the CND is a rotating torus-like molecular gas around Sgr A$^\ast$ \citep[e.g. ][]{Guesten}.  However, the existence of the components being out of  the rotation law indicates that the CND has more complicated structures. The relation between the ionized gas and molecular gas components will be discussed in detail with ALMA molecular line data in another paper.  
 
\begin{figure}
\figurenum{10}
\begin{center}
\includegraphics[width=15cm]{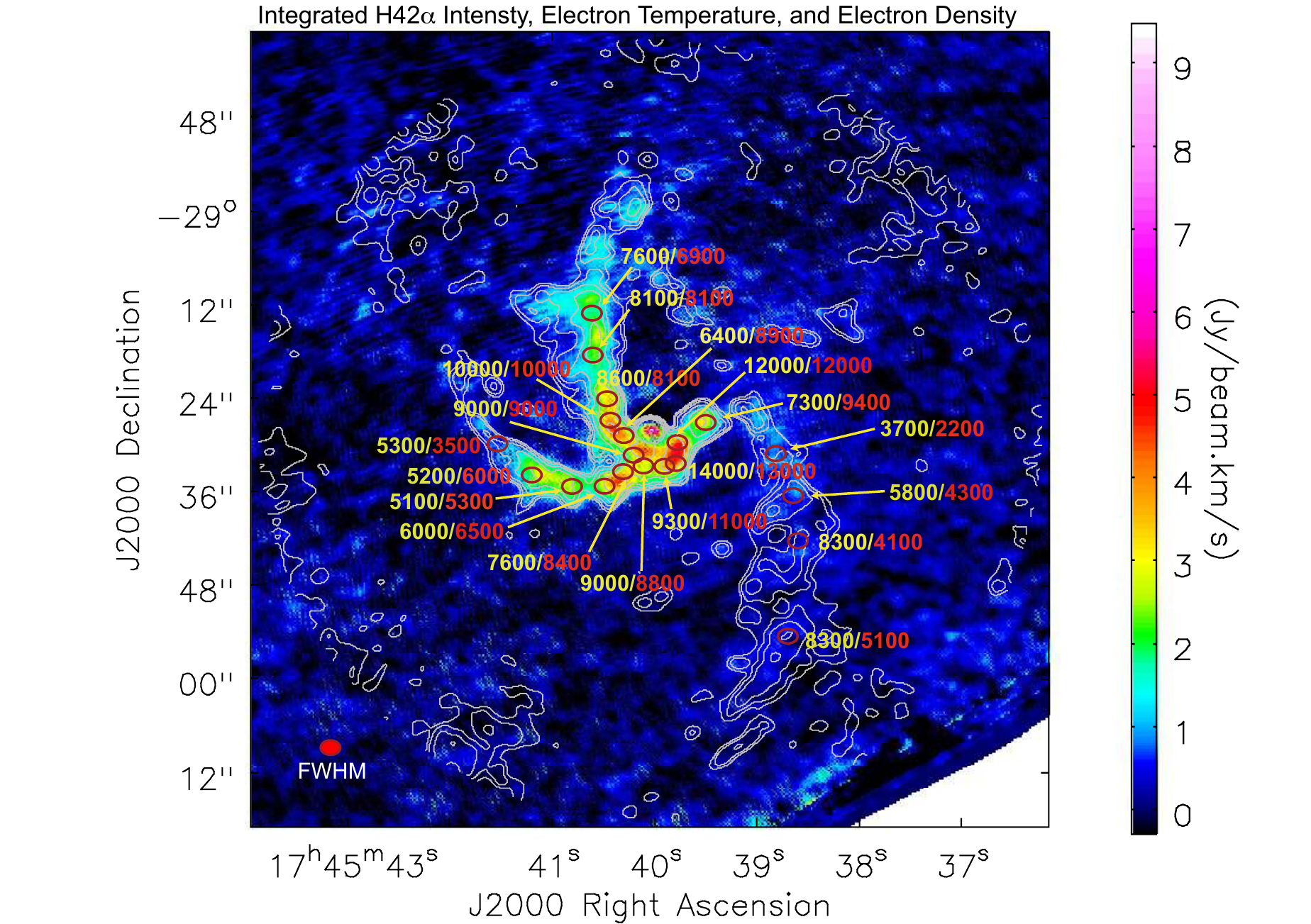}
\caption{Integrated intensity map of the ``Galactic Center Mini-spiral (GCMS)"  in the H42$\alpha$ recombination line (pseudo color) and the electron temperatures in K (numbers in yellow) and the electron densities in cm$^{-3}$ (numbers in red) with the estimated positions indicated by the open ovals. The velocity range of the integrated intensity map is from $V_{\mathrm{LSR}}=-400$ to $+320$ km s$^{-1}$. The angular resolution is $1.9\arcsec \times 1.3\arcsec$($PA=84^\circ$) which is shown on the lower-left corner as a filled oval.  The contours in the figure show the continuum emission of the GCMS at 100 GHz for comparison \citep{Tsuboi2016}.}
\end{center}
\end{figure}
\subsection{Electron Temperature and Electron Density in the Galactic Center Mini-spiral}
\subsubsection{Electron Temperature}
Figure 10 shows the integrated intensity map  of the H42$\alpha$ recombination line. The velocity range is from $V_{\mathrm{LSR}}=-400$ to $+320$ km s$^{-1}$.  The contours in the figure show the continuum flux density at 100 GHz for comparison  \citep{Tsuboi2016}.
The LTE electron temperature, $T^\ast_{\mathrm e}$, in the sub-structures of the GCMS is estimated from the line-continuum flux density ratio shown in figure 10, $S_{\mathrm{line}}/S_{\mathrm{cont}}$, and the observed FWHM velocity width, $\Delta V_{\mathrm{FWHM}}$, assuming that the line and continuum emissions are optically thin. 
The well-known formula of the LTE electron temperature is given by
\begin{equation}
\label{1}
T^\ast_{\mathrm e}[\mathrm K]=\left[\frac{6.985\times10^3}{a(\nu, T^\ast_{\mathrm e})}\Big(\frac{\nu}{\mathrm{GHz}}\Big)^{1.1}
\frac{1}{1+\frac{N(\mathrm{He^+})}{N(\mathrm{H^+})}}
\frac{S_{\mathrm{cont}}(\nu)}
{\int S_{\mathrm{line}}\Big(\frac{dV}{\mathrm{km~s}^{-1}}\Big)}
\right]^{\frac{1}{1.15}}.
\end{equation}
The correction factor, $a(\nu, T^\ast_{\mathrm e})$, for $\nu=86$ GHz and $T^\ast_{\mathrm e}=4\times10^3-1.5\times10^4$ K is $0.822-0.942$ \citep{Mezger}.  We assume that the number ratio of He$^+$ to H$^+$ is $\frac{N(\mathrm{He^+})}{N(\mathrm{H^+})}=0.09$, a typical value for the Orion A HII region \citep[e.g.][]{Rubin}. This is consistent with the non-detection of the He42$\alpha$ recombination line mentioned in section 2.
The LTE electron temperature is obtained by iteratively solving the formula for $T^\ast_{\mathrm e}$. 

As shown in Figure 2 and Figure 3, there are multi-velocity components on the line-of-sight at many positions of the GCMS, and thus it is difficult to derive the electron temperature at such positions. We select the positions where one velocity component is stronger than the other ones by a factor of 4 or more  except for the connecting points among the ionized gas streamers. The derived electron temperatures are also shown in figure 9 (numbers in yellow). These electron temperatures are consistent with typical values in  the HII regions in the Galaxy. The typical uncertainty is estimated to be as large as 15\% of the derived value. These electron temperatures are summarized in Table 1.

 The electron temperatures estimated in the NA are in the range of $T^\ast_{\mathrm e}=(6.4-14.0)\times10^3$ K. The electron temperature of the component at $\alpha=17^h45^m39.8^s$ $\delta=-29^\circ00\arcmin33.4\arcsec$ reaches to the maximum value of $T^\ast_{\mathrm e}=(14.0\pm2.1)\times10^3$ K. 
The electron temperature  in the NA probably increases as approaching Sgr A$^\ast$. The tendency is consistent with  that estimated from previous VLA and SMA observations \citep{Zhao2010}. Because the ionization is originated by UV emission from the Central Cluster around Sgr A$^\ast$, the tendency of the electron temperature may be caused by approaching  Sgr A$^\ast$ along the Keplerian orbit \citep{Zhao2009}. 
On the other hand, the electron temperatures estimated in the EA are lower than those in the NA. They are in the range of $T^\ast_{\mathrm e}=(5.3-6.0)\times10^3$ K.  Although they are fairly monotonous, the highest electron temperature in the EA seems to be also located nearest to Sgr A$^\ast$ on the Keplerian orbit. 

The electron temperatures estimated in the WA are in the range of $T^\ast_{\mathrm e}=(3.7-8.3)\times10^3$ K.  The electron temperatures in the southern half  of the WA may be higher than those in the northern half, and the tendency is different from those seen in the NA and EA. However, note that the highest value in the southernmost part of the WA has the largest uncertainty because the line intensity at the position is very weak as shown in Figure 10.

The electron temperatures at  the west and east ends of the Bar are estimated to be  $T^\ast_{\mathrm e}=7.3\times10^3$ and $7.6\times10^3$ K, respectively.  
At the east end of the Bar, the spectrum contaminates with the component which belongs to the EA.  Because the electron temperature in the EA is as low as $T^\ast_{\mathrm e}\sim6\times10^3$ K as mentioned above, the estimated electron temperature should be fairly lower than the true electron temperatures of the Bar. 

\subsubsection{Electron Density}
The  electron density, $n_{\mathrm e}$, in the sub-structures of the GCMS is estimated from the continuum brightness temperature, $T_{\mathrm B}=1.22\times10^6\Big(\frac{\theta_{{\mathrm maj}}\times\theta_{{\mathrm min}}}{{\mathrm arcsec}^2} \Big)^{-1}\nu^{-2}S_{\mathrm{cont}}$, and  the electron temperature, $T^\ast_{\mathrm e}$, shown in figure 10 and the path length of the ionized gas, $L$, assuming that the continuum emission is optically thin. 
The well-known formula of the electron density is given by
\begin{equation}
\label{2}
n_{\mathrm e}[{\mathrm cm}^{-3}]=\left[\frac{T_{\mathrm B}T_{\mathrm e}^{\ast 0.35}\Big(\frac{\nu}{\mathrm{GHz}}\Big)^{2.1}}{8.235\times10^{-2}\alpha(\nu,T)\Big(\frac{L}{\mathrm{pc}}\Big)}\right]^{0.5}
\end{equation}
\citep{Altenhoff}.
We also assumed here that the $L$ is equal to the observed width of the boundary (see Figures 1 and 2)
and  $n_e$ is constant over the path. 

The derived electron densities are also shown in figure 10 (numbers in red). Although these electron  densities are somewhat  lower than those toward IRS sources in previous observations \citep{Zhao2010}, these are consistent with typical values in  the HII regions in the Galaxy. The typical uncertainty is estimated to be as large as 30\% of the derived value. These electron densities are also summarized in Table 1.

The electron density in the NA appears to increase from $n_{\mathrm e}=7\times 10^3  ~{\mathrm cm}^{-3}$ to $n_{\mathrm e}=13\times 10^3 ~ {\mathrm cm}^{-3}$ with approaching Sgr A$^\ast$. Although a similar tendency is probably seen in the EA, it is not clear in the WA. As mentioned previously, the electron temperature also increases with approaching Sgr A$^\ast$. 
Then the thermal pressure, $n_{\mathrm e}kT_{\mathrm e}$,  of the ionized gas in the inner region of the NA becomes three times or more larger than those in the outer region.  
If there is no confinement by any external pressure as discussed previously, we cannot explain that the electron density and temperature increase without the lateral expansion of the gas streamers.


\begin{table}
  \caption{Electron Temperature and Electron Density in the Galactic Center Mini-spiral. }
  \label{tab:first}
  \begin{center}
    \begin{tabular}{cccccccc}
    \hline
R.A.(J2000) & Dec.(J2000) &$V_{\mathrm{LSR}}^{~~~1}$&$\int S_{\mathrm{line}}(\mathrm{H}42\alpha)dV $& $S_{\mathrm{cont}}(86\mathrm{GHz})^2$  &$T_{\mathrm{e}}^{\ast ~3}$ &$n_{\mathrm{e}}^{~4}$&remark\\
                        &              & [km s$^{-1}$]& [Jy/beam$\cdot$km/s] & [Jy/beam] & [K]&[cm$^{-3}$]&\\
\hline

$17^h45^m39.9^s$ &$-29^\circ00\arcmin30.0\arcsec$&-260/-40&2.503&$10.25\times10^{-2}$ & 12000&$1.2\times10^{4}$&NA/Bar \\
$17^h45^m39.8^s$ &$-29^\circ00\arcmin33.4\arcsec$&-260 &1.645&$10.55\times10^{-2}$ & 14000&$1.3\times10^{4}$&NA \\
$17^h45^m39.9^s$ &$-29^\circ00\arcmin32.9\arcsec$&-240 &2.298&$8.868\times10^{-2}$ & 9300&$1.1\times10^{4}$&NA \\
$17^h45^m40.1^s$ &$-29^\circ00\arcmin32.9\arcsec$&-140&1.592&$5.929\times10^{-2}$ & 9000&$8.8\times10^{3}$&NA \\
$17^h45^m40.2^s$ &$-29^\circ00\arcmin31.6\arcsec$&-100&1.664&$6.210\times10^{-2}$ & 9000&$9.0\times10^{3}$&NA \\
$17^h45^m40.4^s$ &$-29^\circ00\arcmin29.0\arcsec$&-20&2.860&$6.913\times10^{-2}$ & 6400&$8.9\times10^{3}$&NA \\
$17^h45^m40.5^s$ &$-29^\circ00\arcmin27.4\arcsec$&20&1.799&$7.598\times10^{-2}$ & 10000&$1.0\times10^{4}$&NA \\
$17^h45^m40.5^s$ &$-29^\circ00\arcmin24.2\arcsec$&60&2.003&$5.082\times10^{-2}$ & 8600&$8.1\times10^{3}$&NA\\
$17^h45^m40.6^s$ &$-29^\circ00\arcmin18.6\arcsec$&120&1.558&$5.183\times10^{-2}$ & 8100&$8.1\times10^{3}$&NA \\
$17^h45^m40.6^s$ &$-29^\circ00\arcmin13.9\arcsec$&120&1.304&$3.910\times10^{-2}$ & 7600&$6.9\times10^{3}$&NA \\

$17^h45^m39.5^s$ &$-29^\circ00\arcmin27.2\arcsec$&-160&2.552&$7.341\times10^{-2}$ & 7300&$9.4\times10^{3}$&Bar \\
$17^h45^m40.3^s$ &$-29^\circ00\arcmin33.2\arcsec$&150/300&1.900&$5.740\times10^{-2}$ & 7600&$8.4\times10^{3}$&EA/Bar \\

$17^h45^m40.5^s$ &$-29^\circ00\arcmin35.4\arcsec$&160&1.709&$3.787\times10^{-2}$ & 6000&$6.5\times10^{3}$&EA \\
$17^h45^m40.8^s$ &$-29^\circ00\arcmin35.6\arcsec$&160&1.447&$2.638\times10^{-2}$ & 5100&$5.3\times10^{3}$&EA\\
$17^h45^m41.2^s$ &$-29^\circ00\arcmin34.2\arcsec$&140&1.776&$3.330\times10^{-2}$ & 5200&$6.0\times10^{3}$&EA \\
$17^h45^m41.4^s$ &$-29^\circ00\arcmin30.2\arcsec$&140&0.735&$1.120\times10^{-2}$ & 5300&$3.5\times10^{3}$&EA \\

\hline
    \end{tabular}
 \end{center}
 $^1$ The LSR velocity at the peak of the main component.
$^2$ We assumed $S_{\mathrm{cont}}(86\mathrm{GHz})=S_{\mathrm{cont}}(100\mathrm{GHz})(86/100)^{-0.1}\Big(\frac{\Omega_{\mathrm{beam}}(86)}{\Omega_{\mathrm{beam}}(100)}\Big)$. $S_{\mathrm{cont}}(100\mathrm{GHz})$ is calculated from  the previous observation of the ``Galactic Center Mini-spiral" at 100 GHz \citep{Tsuboi2016}. $^3$  The typical uncertainty is estimated to be as large as 15\% of the derived value. $^4$ The typical uncertainty is estimated to be as large as 30\% of the derived value.\\
\clearpage
\end{table}
\begin{table}
\addtocounter{table}{-1}	
  \caption{Continued. }
  \label{tab:first}
  \begin{center}
 \begin{tabular}{cccccccc}
    \hline
R.A.(J2000) & Dec.(J2000) &$V_{\mathrm{LSR}}$&$\int S_{\mathrm{line}}(\mathrm{H}42\alpha)dV $& $S_{\mathrm{cont}}(86\mathrm{GHz})$  &$T_{\mathrm{e}}^{\ast}$ &$n_{\mathrm{e}}$&remark\\
                        &              & [km s$^{-1}$]& [Jy/beam$\cdot$km/s] & [Jy/beam] & [K]&[cm$^{-3}$]&\\
                        \hline

$17^h45^m38.9^s$ &$-29^\circ00\arcmin31.6\arcsec$&-20&0.440&$0.529\times10^{-2}$ & 3700&$2.2\times10^{3}$&WA \\
$17^h45^m38.6^s$ &$-29^\circ00\arcmin36.0\arcsec$&-30&0.768&$1.655\times10^{-2}$ & 5800&$4.3\times10^{3}$&WA \\
$17^h45^m38.6^s$ &$-29^\circ00\arcmin42.4\arcsec$&-40&0.403&$1.348\times10^{-2}$ &8300&$4.1\times10^{3}$&WA \\
$17^h45^m38.7^s$ &$-29^\circ00\arcmin54.5\arcsec$&-100&0.599&$2.022\times10^{-2}$ &8300&$5.1\times10^{3}$&WA \\

\hline
    \end{tabular}
 \end{center}
\clearpage
\end{table}

\subsection{Protostar Candidates in the Galactic Center Mini-spiral}
A mm-wave continuum observation with JVLA found several compact components near Sgr A$^\ast$, which have ionized half-shell-like structures facing Sgr A$^\ast$.  These components could be protostar candidates with low mass; their surfaces facing  Sgr A$^\ast$ are rapidly photoevaporated by strong Lyman continuum emitting from the Central Cluster around Sgr A$^\ast$ \citep{Yusef-Zadeh2015}. However, the physical relation between these  candidates of low-mass protostar and the ionized gas streamers of the GCMS has not been clarified yet. 

The upper panel in Figure 11 shows an integrated intensity map for the EA and NEA in the H42$\alpha$ recombination line from $V_{\mathrm{LSR}}=+130$ to $+150$ km s$^{-1}$ (see also Figure 1).  
The NEA component seen in the H42$\alpha$ recombination line is located at the southwestern tip of the NEA traced by the 100 GHz continuum. Moreover, the component seems to be accompanied by a faint component seen in the CS emission line \citep{Tsuboi2016b}. 

The lower panel in Figure 11 shows the comparison between the ionized gas streamer and  the photoevaporating low-mass protostar candidates. The black-and-white image shows the continuum emission at 34 GHz observed by JVLA \citep{Yusef-Zadeh2015}. The protostar candidates detected in the 34-GHz continuum map are concentrated exclusively in the ionized gas streamer of the NEA traced by the H42$\alpha$ recombination line.  This good positional correlation indicates that the protostar candidates are physically associated with the NEA. 

As mentioned in Introduction, the tidal force of Sgr A$^\star$ and the strong Lyman continuum radiation from the Central cluster must suppress the star-formation activity near Sgr A$^\ast$. 
The positional correlation in Figure 11 suggests that the protostar candidates had formed in the NEA when it was located in the outer region and they were brought to near Sgr A$^\ast$ according as the streamer fell.  If a star-forming molecular cloud fall from an outer region and supplies young stars to the vicinity of Sgr A$^\star$,  the obstacles to the star-formation activity near Sgr A$^\ast$ should be overcome.
\begin{figure}
\figurenum{11}
\begin{center}
\includegraphics[width=12cm]{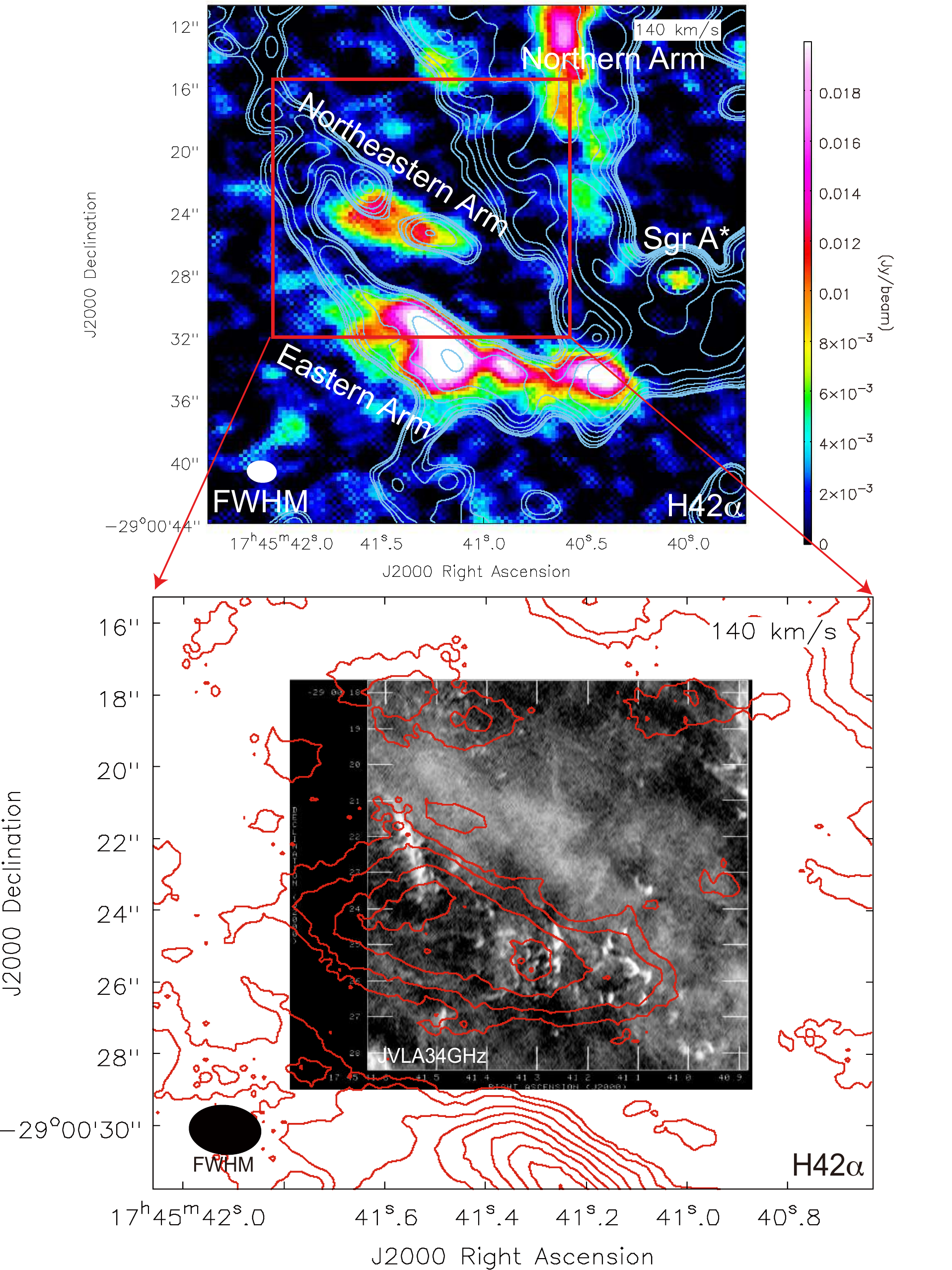}
\caption{{\bf upper panel}:Integrated intensity map of the H42$\alpha$ recombination line (pseudo color). The velocity range of the map is from $V_{\mathrm{LSR}}=+130$ to $+150$ km s$^{-1}$. The angular resolution is $1.9\arcsec \times 1.3\arcsec$($PA=84^\circ$) which is shown on the lower-left corner as a white filled oval.  The contours in the figure show the continuum emission of the ``Galactic Center Mini-spiral (GCMS)" at 100 GHz for comparison \citep{Tsuboi2016}. {\bf lower panel}:Integrated intensity map of  H42$\alpha$ recombination line (contours) of a rectangle area shown in the upper panel. The velocity range of the integration is the same as in the upper panel. The contour levels are 1.9, 3.8, 7.5, 11, 15, 19, 23, 26, and 30 mJy beam$^{-1}$. The angular resolution same as in the upper panel is shown on the lower-left corner as a black filled oval. The black-and-white image is the continuum emission at 34 GHz observed by JVLA \citep{Yusef-Zadeh2015}. }
\end{center}
\end{figure}

\section{Summary}
 We have performed the observation of the GCMS in the H42$\alpha$ recombination line as a part of  the first large-scale mosaic observation in the Sgr A complex using the Atacama Millimeter/sub-millimeter Array (ALMA). 
We found that the ionized gas streamers of the NA and EA in their outer regions somewhat deviate  from the Keplerian orbits which were derived previously from the trajectories in the inner regions. 
In addition, we found that the ionized gas streamer corresponding to the Bar has an independent Keplerian orbit with an eccentricity of $e\sim0.8$. The periastron is probably located within the Bondi accretion radius of Sgr A$^{\ast}$. 
We estimated the electron temperature and  electron density in the ionized gas streamers. The electron temperatures are in the range of $T^\ast_{\mathrm e}=(4-14)\times10^3$ K. We confirmed the previously claimed tendency that the electron temperatures increase toward Sgr A$^\ast$. 
We found that the electron density in the NA and EA also increases with approaching Sgr A* without the lateral expansion of the streamers.
This suggests that there is some external pressure around the GCMS. If the pressure is caused by the ambient ionized gas, the ambient gas may affect the orbits of the ionized gas streamers by Ram pressure.
There is a good positional correlation between the protostar candidates detected by JVLA at 34 GHz and the ionized gas streamer NEA found in the north of the EA  by the H42$\alpha$ recombination line observation. This suggests that the protostar candidates had already formed in the streamer and they were brought to near Sgr A$^\ast$ according as the streamer fell. 

\acknowledgments
 The National Radio Astronomy Observatory (NRAO) is a facility of the National Science Foundation (NSF) operated under cooperative agreement by Associated Universities, Inc (AUI). This paper makes use of the following ALMA data:ADS/JAO.ALMA\#2012.1.00080.S.  ALMA is a partnership of ESO (representing its member states), NSF (USA) and NINS (Japan), together with NRC(Canada), NSC and ASIAA (Taiwan), and KASI (Republic of Korea), in cooperation with the Republic of Chile. The Joint ALMA Observatory is operated by ESO, AUI/NRAO and NAOJ. 
This work is supported in part by the Grant-in-Aid from the Ministry of Eduction, Sports, Science and Technology (MEXT) of Japan, No.16K05308.

\vspace{5mm} 
\facilities{ALMA}
\software{CASA}

\clearpage
\section{Appendix; Position-Velocity Diagram of Keplerian Orbits with High Eccentricity}
As well known, the position of a test particle on a Keplerian elliptical orbit around a heavy mass, $M$, is shown by
$$ x=a(\cos u-e), ~~~~y=a\sqrt{1-e^2}\sin u,$$
where $a$, $e$, and $u$ are the semi-major axis, the eccentricity, and the eccentric anomaly, respectively. On the other hand, the orbital velocity components of the test particle are given by
$$ V_x=-\frac{na\sin u}{1-e\cos u},~~~~V_y=\frac{na\sqrt{1-e^2}\cos u}{1-e\cos u},$$
where $n$ is the mean motion, $n=\sqrt{\frac{GM}{a^3}}$.  If $u$ is used as a parameter, the relations between  $x$ and $ V_y$ and between  $y$ and $ V_x$ 
are easily given. When the orbit is observed from a very far place,  the relation is called the position-velocity diagram of the orbit. The direction of the line-of-sight is shown by position angle, $PA$, which is the angle between the line-of-sight and the major axis of the orbit. Then the position-velocity diagram is given by the following equations;
$$P=-x(0)\sin PA+y(0)\cos PA, ~~~~V=V_x(0)\cos PA+V_y(0)\sin PA.$$
Figure 6 shows the position-velocity diagrams of the Keplerian oval orbits with $e=0.5-0.95$ on the various position angles of the line-of-sight. In the calculation, we assumed that $na=1$. The relation between the line-of-sight and the elliptical orbit is shown in lower-left corner of each panel.  The measured position-velocity diagram shown in figure 4 is the convolution of the orbit motion shown in figure 8 with the internal motions which are turbulent and thermal velocities for the case of ionized gas.
\begin{figure}
\figurenum{12}
\begin{center}
\includegraphics[width=15cm]{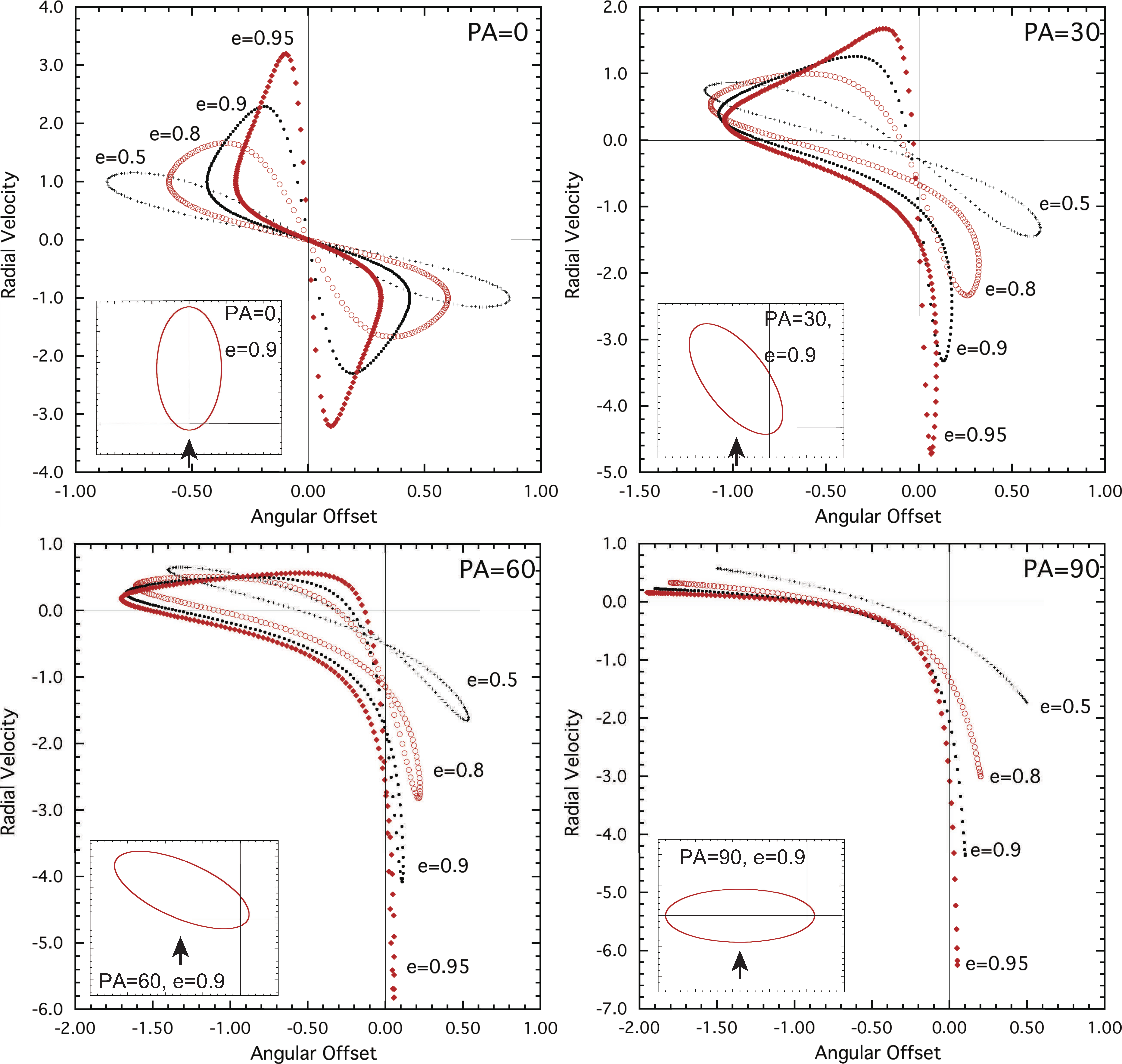}
\caption{The position-velocity diagrams of the Keplerian oval orbits with $e=0.5-0.95$ various position angles of the line-of-sight. In the calculation, we assumed that $na=1$.  The relation between the line-of-sight (arrow) and the elliptical orbit is shown in the lower-left corner of each panel.  }
\end{center}
\end{figure}

\end{document}